\RequirePackage{lineno}
\documentclass[aps,prd,10pt,twocolumn,superscriptaddress,showpacs,longbibliography,nofootinbib]{revtex4-1}
\usepackage{graphicx}
\usepackage{amsfonts,amsmath,amssymb,bm,bbm}
\usepackage{color}
\usepackage{enumitem}
\usepackage{url}  

\usepackage{latexsym,epsfig,multirow,comment,appendix,feyn,slashed,afterpage,makecell} 

\usepackage{aas_macros}
\usepackage{mathrsfs}
\usepackage[dvipsnames, usenames]{xcolor}
\usepackage[normalem]{ulem}
\usepackage[pdftex]{hyperref}
\usepackage[capitalize]{cleveref}

\definecolor{romared}{RGB}{142,0,28}
\usepackage{hyperref}
\hypersetup{
    pdfnewwindow=true,      
    colorlinks=true,       
    linkcolor=romared,          
    citecolor=romared,        
    filecolor=romared,      
    urlcolor=romared        
}

\def\bi{\begin{itemize}[noitemsep,leftmargin=*]
\setlength\itemsep{1em}
        }
\def\ei{\end{itemize}}

\newcommand{\pd}{\partial}  
\newcounter{ichi}
\newcounter{ni}
\newcounter{san}
\newcounter{yon}
\newcommand{\orcid}[1]{\href{https://orcid.org/#1}{\includegraphics[width=10pt]{orcid.pdf}}}
\newcommand{\github}[1]{\href{https://github.com/#1}{\includegraphics[width=10pt]{github.pdf}}}

\begin{document}


\title{Interacting supernovae as high-energy multimessenger transients}
\author{Kohta Murase}
\affiliation{Department of Physics; Department of Astronomy \& Astrophysics; Center for Multimessenger Astrophysics, Institute for Gravitation and the Cosmos, The Pennsylvania State University, University Park, Pennsylvania 16802, USA}
\affiliation{School of Natural Sciences, Institute for Advanced Study, Princeton, New Jersey 08540, USA}
\affiliation{Center for Gravitational Physics and Quantum Information, Yukawa Institute for Theoretical Physics, Kyoto, Kyoto, 606-8502, Japan}

\date{submitted 24 December 2023}

\begin{abstract}
Multiwavelength observations have revealed that dense, confined circumstellar material (CCSM) commonly exists in the vicinity of supernova (SN) progenitors, suggesting enhanced mass losses years to centuries before their core collapse. Interacting SNe, which are powered or aided by interaction with the CCSM, are considered to be promising high-energy multimessenger transient sources. We present detailed results of broadband electromagnetic emission, following the time-dependent model proposed in the previous work on high-energy SN neutrinos [K. Murase, New prospects for detecting high-energy neutrinos from nearby supernovae, Phys. Rev. D 97, 081301(R) (2018)]. We investigate electromagnetic cascades in the presence of Coulomb losses, including inverse-Compton and synchrotron components that significantly contribute to MeV and high-frequency radio bands, respectively. We also discuss the application to SN 2023ixf. 
\end{abstract}

\maketitle

\section{Introduction}\label{sec:intro}
Multiwavelength observations of supernovae (SNe) in optical, radio, and x-ray bands have provided cumulative evidences for the existence of dense circumstellar material (CSM) surrounding their progenitors~\cite{Smith:2016dnb,Maeda:2022gyd}, which has led to a paradigm shift in the stellar evolution study~\cite{Smith:2014txa}. 
A good fraction of hydrogen-rich (Type II) SNe show signatures of confined CSM (CCSM), which include narrow emission lines~\cite{Khazov:2015nzj,Yaron:2017umb,BG20,Terreran:2021hfc,Bruch:2020jcr,Bruch:2022aqd} and optical light curves at early times~\cite{Morozova:2016efp,Das:2017wen,Forster:2018mib}. 
The radiative output of interaction-powered SNe, which are often classified as Type IIn SNe, is dominated by the shock interaction between the supernova (SN) ejecta and the CSM~\cite{Smith:2007cb,Immler:2007mk,Miller:2008jy}). 
Variable activities of progenitors or even outbursts that occur months to years before explosion are also observed~\cite{Ofek:2013mea,Ofek:2014ifa,Margutti:2013pfa,Jacobson-Galan:2021pki}. 
The interaction with such CCSM can be important even in Type Ibc SNe, as evidenced from transrelativistic SNe~\cite{Campana:2006qe,Nakar:2011mq}, some fast blue optical transients like AT 2018cow~\cite{Margutti:2018rri,Fox:2019zmd}, and interacting SNe like SN 2014C~\cite{Margutti:2016wyh}. 
The diversity of such ``interacting SNe''\footnote{Interacting SNe refer to any type of SNe experiencing shock interaction with CSM, which include not only SNe IIn but also SNe II with CCSM.}, whether the optical emission is mainly powered by the CSM interaction or not, suggests the importance of understanding the final phase in the evolution of massive stars~\cite{Woosley:2007qp,Moriya:2011aa,Quataert:2012pg,Chevalier:2012ba,Fuller:2017tgs}, and the progenitors of core-collapse SNe may commonly experience significant mass losses ranging from $\sim10^{-3}$ to $\sim1~M_\odot~{\rm yr}^{-1}$ that occur $\sim0.1-100$~yr before the explosion~\cite{Smith:2014txa,Moriya:2014cua,Tsuna:2021zzx}. This picture has also been supported by late-time radio observations of core-collapse SNe~\cite{Stroh:2021yya}. 
The recent discovery of SN 2023ixf at a distance of $d\sim7$~Mpc provided another golden example of CCSM-interacting SNe II with various signatures at different wavelengths~\cite{Yamanaka:2023gbr,Jacobson-Galan:2023ohh,Bostroem:2023dvn,Teja:2023hcm,Hiramatsu:2023inb,Grefenstette:2023dka,Chandra:2023kda,Zimmerman:2023mls,Li:2023vux}.

The collision between SN ejecta and dense CSM creates a rich tapestry of physical processes and will lead to various high-energy signatures. Large CSM masses inevitably cause efficient dissipation of the kinetic energy of the SN ejecta via shocks~\cite{Chevalier:2016hzo}. The formation of collisionless shocks will be accompanied by the onset of diffusive shock acceleration (DSA) of cosmic rays (CRs) and resulting nonthermal emission~\cite{Murase:2010cu,Katz:2012zzk}. Murase et al.~\cite{Murase:2010cu} proposed that interaction-powered SNe like Type IIn SNe are promising sources of high-energy neutrinos and gamma rays. Murase~\cite{Murase:2017pfe} (hereafter M18) first investigated high-energy neutrino emission from various types of SNe considering interaction with CCSM (including ordinary SNe II-P) and showed that Galactic SNe are promising multienergy neutrino transients for neutrino detectors such as IceCube and Super-Kamiokande. Detection of high-energy SN neutrinos has been of interest for studying particle physics~\cite{Wen:2023ijf}.   
Hadronic gamma-ray emission from interacting SNe has also been investigated both theoretically~\cite{Murase:2010cu,Kashiyama:2012zn,Murase:2018okz,Cristofari:2020pzg,Cristofari:2022low,Sarmah:2022vra,Brose:2022mbx,Sarmah:2023sds,Guarini:2023rnd,Sarmah:2023xrm} and observationally~\cite{Fermi-LAT:2015cpa}, and analyses with the {\it Fermi} data enable us to obtain meaningful constraints on the CR acceleration from SN 2010jl~\cite{Murase:2018okz} and find the possible signal from bright SNe~\cite{Yuan:2017oqu,Xi:2020mif,Chen:2023fsm}. 
This CCSM interaction scenario does not rely on an additional central engine source, and it is different from the scenario assuming a pulsar embedded in SN ejecta to be a CR accelerator~\cite{BG87,Gaisser:1987wm,Murase:2009pg,Kotera:2013yaa,Murase:2014bfa}. 
Interacting SNe can be super-PeVatrons~\cite{Murase:2013kda,Marcowith:2018ifh,Inoue:2021bjx,Diesing:2023ldd}, which may also contribute to Galactic CRs~\cite{Sveshnikova:2003sa,Zirakashvili:2015mua,Kawanaka:2021rbe}. 
Revealing high-energy nonthermal emission from such interacting supernovae will not only expand our understanding of the late stages of stellar evolution but also shed light on the properties of particle acceleration. 

In this work, we investigate high-energy nonthermal emission from various classes of SNe considering early CSM interactions lasting from days to months, and provide details of numerical calculations with the Astrophysical Multimessenger Emission Simulator ({\sc AMES}). In particular, following M18 that focuses on high-energy SN neutrino afterglows, we present the results on broadband electromagnetic emission, taking into account synchrotron and inverse-Compton (IC) processes and accompanied electromagnetic cascades. The results will be useful for multimessenger modeling of nearby SNe and distant SNe that could be found by neutrino-triggered followup observations. 
Throughout this work, we use $Q\equiv10^{x}Q_x$ in CGS units.

\begin{table*}[t]
\begin{center}
\caption{CSM and SN ejecta parameters for various types of SNe considered in this work and M18. Kinetic energy and mass of SN ejecta are set to ${\mathcal E}_{\rm ej}=10^{51}$~erg and $M_{\rm ej}=10~M_\odot$, respectively. 
\label{tab:CCSM}
}
\scalebox{1.0}{
\begin{tabular}{|c||c|c|c||c||c|c||c|}
\hline Class & $D_*$ & $\dot{M}_{w}$ [$M_{\odot}$~yr$^{-1}$] & $V_{w}$ [km~s$^{-1}$] & $R_{\rm cs}$ [cm] & $R_*$ [cm] & $\delta$ & $t_{\nu \rm max}$ [s]\\
\hline II (CCSM) & ${10}^{-2}-1$ & ${10}^{-3}-{10}^{-1}$ & $100$ & $(0.4-1.0)\times{10}^{15}$ & $6\times{10}^{13}$ & 12 & ${10}^{5.8}-{10}^{6.5}$ \\
\hline IIn & ${10}^{-2}-1$ & ${10}^{-3}-{10}^{-1}$ &  $100$ & ${10}^{16}$ & ${10}^{13}$ & 10 & ${10}^{5.8}-{10}^{7.5}$\\
\hline II-P & $1.34\times{10}^{-4}$ & $2\times{10}^{-6}$ & $15$ & -  & $6\times{10}^{13}$ & 12 & ${10}^{5.4}$\\
\hline II-L/IIb & ${10}^{-3}$ & $3\times{10}^{-5}$ &  $30$ & - & $6\times{10}^{12}$ & 12 & ${10}^{4.6}$\\
\hline Ibc & ${10}^{-5}$ & ${10}^{-5}$ & $1000$ & - & $3\times{10}^{11}$ & 10 & ${10}^{3.8}$\\
\hline
\end{tabular}
}
\end{center}
\end{table*}

\section{Physical Model}\label{sec:physics}
\subsection{Dynamics}
Let us consider SN ejecta with an ejecta mass of $M_{\rm ej}=10~M_\odot M_{\rm ej,1}$, which interact with a CCSM (that can also be an extended stellar envelope) with a density profile of 
\begin{equation}
\varrho_{\rm cs}=D R_{\rm cs}^{-2} {\left(\frac{r}{R_{\rm cs}}\right)}^{-w},
\end{equation}
where $R_{\rm cs}$ is the outer radius of the CCSM and $w$ is a CSM density slope. We use a wind profile (i.e., $w=2$), which is reasonable and sufficient for the purpose of this work, although a shallower profile has been discussed both theoretically~\cite{Tsuna:2021bvq} and observationally~\cite{Stroh:2021yya}.  
In the wind case, the CSM parameter is written as
\begin{equation}
D\equiv 5\times{10}^{16}~{\rm g~{\rm cm}^{-1}}~D_*=\frac{\dot{M}_{w}}{4\pi f_{\Omega} V_w},
\end{equation}
where ${\dot{M}_{w}}$ is the wind mass-loss rate, $V_w$ is the wind velocity, and $f_\Omega\equiv \Delta\Omega/(4\pi)$ is the covering factor of the CSM, which can be lower than unity if the CSM is aspherical and/or clumpy. 
For example, early flash spectroscopy for SN 2013fs indicates $D_*\sim{10}^{-2}$ and $R_{\rm cs}\sim4\times10^{14}$~cm~\cite{Yaron:2017umb}. Modeling of SN 2020tlf (II-P) suggests $D_*\sim0.6$ and $R_{\rm cs}\sim10^{15}$~cm~\cite{Chugai:2022fcz} (see also Ref.~\cite{Jacobson-Galan:2021pki}). As an example of SNe IIn, observations of SN 2010jl infer $D_*\sim6$ and $R_{\rm cs}\sim10^{16}$~cm~\cite{Ofek:2013afa,Fransson:2013qya}. Ref.~\cite{BG20} shows that interacting SNe have a range of parameters from $D_*\sim{10}^{-2}-1$. 
The number density of nucleons is $n_N=\varrho_{\rm cs}/m_H$ and that in the shocked region $n'_N$ is larger by the compression ratio.  
The total CSM mass ($M_{\rm cs}$) is also obtained by performing the volume integral. 
CSM and SN ejecta parameters used in this work are shown in Table~\ref{tab:CCSM}. Note that although we primarily consider Type II SNe for CCSM-interacting SNe, some SNe Ibc such as SN 2010bh, SN 2014C and AT 2018cow are accompanied by CCSM, which enables us to expect high-energy neutrino and gamma-ray emission~\cite{Kashiyama:2012zn,Murase:2018okz}.

A faster component of the SN ejecta is decelerated earlier, and the forward shock evolution is obtained by solving the equation for the conservation of momentum~\cite{Chevalier:1982,Nadezhin:1985}. In the thin shell approximation, the radius ($R_s$) and velocity ($V_s$) of the shell are determined by~\cite{Moriya:2013hka,Li:2023vux}
\begin{eqnarray}
M_{s}\frac{dV_{s}}{dt}=4\pi f_\Omega R_s^2[\varrho_{\rm ej}{(V_{\rm ej}-V_s)}^2-\varrho_{\rm cs}{(V_{s}-V_w)}^2],
\label{eq:Rs}
\end{eqnarray}
where $\varrho_{\rm ej}\propto t^{-3}{(r/t)}^{-\delta}$ (where $\delta\geq6.67$) is the the outer ejecta profile, $M_{s}$ is mass of the shell consisting of the shocked SN and CSM, and $V_{\rm ej}$ is the ejecta velocity. We adopt $\delta=12$ ($\delta=10$) for supergiant stars with a radiative envelope (Wolf-Rayet-like compact stars with a convective envelope)~\cite{Matzner:1998mg}, although our conclusions are largely insensitive to this assumption. Typical progenitors of SNe II-P and II-L/IIb are thought to be red supergiants (RSGs) and yellow supergiants, respectively~\cite{Smith:2014txa}.
Practically, a power-law solution of Eq.~(\ref{eq:Rs}), which is obtained with $V_{\rm ej}=R_s/t$ and $V_s\gg V_w$, is valid until the shock radius reaches $R_{\rm s}=V_t t_t$, where $V_t={[10(\delta-5){\mathcal E}_{\rm ej}/3/(\delta-3)/M_{\rm ej}]}^{1/2}$~\cite{Moriya:2013hka}. The deceleration of the whole ejecta would happen if $M_{\rm ej}\lesssim M_{\rm cs}$.

While the solution of Eq.~(\ref{eq:Rs}) is used in our numerical calculations~\cite{Murase:2017pfe}, analytical estimates would be useful for understanding the physics. The shock radius is approximately given by
\begin{equation}
R_{s}\simeq
\left\{
\begin{array}{lr} 
2.4\times{10}^{14}~{\rm cm}~D_{*,-2}^{-1/10}{\mathcal E}_{\rm ej,51}^{9/20}M_{\rm ej,1}^{-7/20}t_{5.5}^{9/10} & \mbox{($\delta=12$)}  \,\,\, \,\,\, \,\,\, \,\,\, \,\,\, \,\,\,\\
1.7\times{10}^{14}~{\rm cm}~D_{*}^{-1/8}{\mathcal E}_{\rm ej,51}^{7/16}M_{\rm ej,1}^{-5/16}t_{5.5}^{7/8} & \mbox{($\delta=10$)}  \,\,\, \,\,\, \,\,\, \,\,\, \,\,\, \,\,\,
\end{array} 
\right., 
\label{RsIIP}
\end{equation}
and the corresponding velocity $V_s=dR_s/dt$ is
\begin{equation}
V_{s}\simeq
\left\{
\begin{array}{lr} 
6.9\times{10}^{8}~{\rm cm}~{\rm s}^{-1}~D_{*,-2}^{-1/10}{\mathcal E}_{\rm ej,51}^{9/20}M_{\rm ej,1}^{-7/20}t_{5.5}^{-1/10}  \\
 \mbox{(for $\delta=12$)}  \,\,\, \,\,\, \,\,\, \,\,\, \,\,\, \,\, \,\,\, \,\,\, \,\,\, \,\,\, \,\,\, \,\,\, \,\,\, \,\,\,  \,\,\, \\
4.6\times{10}^{8}~{\rm cm}~{\rm s}^{-1}~D_{*}^{-1/8}{\mathcal E}_{\rm ej,51}^{7/16}M_{\rm ej,1}^{-5/16}t_{5.5}^{-1/8} \\
 \mbox{(for $\delta=10$)}  \,\,\, \,\,\, \,\,\, \,\,\, \,\,\, \,\, \,\,\, \,\,\, \,\,\, \,\,\, \,\,\, \,\,\, \,\,\, \,\,\,  \,\,\,
\end{array} 
\right..
\label{VsIIP}
\end{equation}

A fraction of the shell kinetic energy that is dissipated at a shock is converted into internal energy, magnetic fields, and CRs. 
The kinetic luminosity, $L_s=2\pi f_{\Omega}\varrho_{\rm cs}V_s^3R_s^2$, is estimated to be
\begin{equation}
L_{s}\simeq
\left\{
\begin{array}{lr} 
1.0\times{10}^{42}~{\rm erg}~{\rm s}^{-1}~f_\Omega D_{*,-2}^{7/10}{\mathcal E}_{\rm ej,51}^{27/20}M_{\rm ej,1}^{-21/20}t_{5.5}^{-3/10} \\
\mbox{(for $\delta=12$)}  \,\,\, \,\,\, \,\,\, \,\,\, \,\,\, \,\, \,\,\, \,\,\, \,\,\, \,\,\, \,\,\, \,\,\, \,\,\, \,\,\,  \,\,\, \\
3.1\times{10}^{43}~{\rm erg}~{\rm s}^{-1}~f_\Omega D_{*}^{5/8}{\mathcal E}_{\rm ej,51}^{21/16}M_{\rm ej,1}^{-16/15}t_{5.5}^{-3/8}  \\
\mbox{(for $\delta=10$)}  \,\,\, \,\,\, \,\,\, \,\,\, \,\,\, \,\, \,\,\, \,\,\, \,\,\, \,\,\, \,\,\, \,\,\, \,\,\, \,\,\,  \,\,\,
\end{array} 
\right..
\label{LdIIP}
\end{equation}
For the demonstrative purpose of this work, it is sufficient to assume $f_\Omega=1$. In reality, the observed flux is related to $L_s$, and uncertainty in $f_\Omega$ degenerates with uncertainties in the other parameters.

\subsection{CR acceleration and secondary production}
It is believed that GeV--PeV CRs originate from SN remnants with an age of $\sim10^3-10^4$~yr (that is comparable to the deceleration time of the whole SN ejecta), and shock interactions with stellar winds in the compact clusters of young massive stars may also be important for CRs around or above the knee energy~\cite{Bykov:2017fik}. Shell-type SN remnants are established as efficient particle accelerators~\cite{Funk:2015ena}, where both ions and electrons are accelerated by the DSA mechanism~\cite{Drury:1983zz,Caprioli:2023orv} that is one of the Fermi acceleration processes~\cite{Fermi:1949ee}. 
However, a shell caused by SN shocks had not been thought as promising high-energy neutrino and gamma-ray emitters during the early phases with an age of $\lesssim0.1-1$~yr after the SN explosion. 
First, most of the SN ejecta freely expands just after the SN explosion, so that energy carried by CRs via DSA is small especially if the shock propagates in the interstellar medium~\cite{BG87}. Alternatively, a pulsar can be invoked as a CR accelerator but the CR acceleration mechanism is highly uncertain~\cite{BG87,Gaisser:1987wm,Murase:2009pg}. 
Second, DSA at a shock inside a star is inefficient while the shock is collisional or radiation mediated~\cite{Weaver:1976,Waxman:2001kt}. This is still the case around shock breakout when the density profile of the material is so steep that the shock is subject to significant radiative acceleration~\cite{Katz:2012zzk,Murase:2018okz,Tsuna:2023bvg}. 
Third, there has been remarkable progress in gamma-ray and neutrino observations. For example, IceCube is sensitive enough to detect high-energy neutrino signals if Betelgeuse explodes even without considering enhanced CSM~\cite{Murase:2017pfe}.    

However, the common existence of CCSM in core-collapse SNe including ordinary SNe II-P has changed prospects for high-energy neutrino and gamma-ray emission, as pointed out by M18. 
Efficient DSA may start once the SN shock leaves a star, for which the condition is given by $R_*=R_s(t_*)$ for $V_s<V_{s,\rm max}$ (where $V_{s,\rm max}$ is the maximum velocity~\cite{Matzner:1998mg}), and we have
\begin{eqnarray}
t_{*}\simeq
\left\{ \begin{array}{ll} 
6.8\times{10}^{4}~{\rm s}~D_{*,-2}^{1/9}M_{\rm ej,1}^{7/18}{\mathcal E}_{\rm ej,51}^{-1/2}R_{*,13.78}^{10/9}\\
\mbox{(for $\delta=12$)} \,\,\, \,\,\, \,\,\, \,\,\, \,\,\, \,\, \,\,\, \,\,\, \,\,\, \,\,\, \,\,\, \,\,\, \,\,\, \,\,\,  \,\,\, \\
1.3\times{10}^{4}~{\rm s}~D_{*}^{1/7}M_{\rm ej,1}^{5/14}{\mathcal E}_{\rm ej,51}^{-1/2}R_{*,13}^{8/7}\\
\mbox{(for $\delta=10$)} \,\,\, \,\,\, \,\,\, \,\,\, \,\,\, \,\, \,\,\, \,\,\, \,\,\, \,\,\, \,\,\, \,\,\, \,\,\, \,\,\,  \,\,\, \\
\end{array} \right..
\end{eqnarray} 
If the CSM is too dense, the shock is initially radiation mediated, in which the shock jump is smeared out by radiation from the downstream and low-energy CRs gain little energy~\cite{Murase:2013kda}. However, the formation of collisionless shocks (mediated by plasma instabilities) is unavoidable especially for a wind or shallower density profile~\cite{Katz:2012zzk,Kashiyama:2012zn,Murase:2013kda}, and the condition is given by $\tau_T\approx \sigma_T{\varrho}_{\rm cs}R_{s}/(\mu_e m_H)\lesssim c/V_s$, where $\tau_T$ is the optical depth to the Thomson scattering with the cross section, $\sigma_T\approx6.7\times{10}^{-25}~{\rm cm}^2$. This coincides with the photon breakout time~\cite{Chevalier:2011ha}, which is. 
\begin{eqnarray}
t_{\rm bo}\simeq5.8\times{10}^{3}~{\rm s}~D_{*,-2}\mu_e^{-1}.
\end{eqnarray}
Considering these two necessary criteria, the onset time of CR acceleration is estimated by~\cite{Murase:2017pfe} 
\begin{equation}
t_{\rm onset}\simeq{\rm max}[t_{\rm bo},t_*].
\end{equation}
For $D_*\lesssim0.1$, which is the case for most SNe II, we expect $t_{\rm onset}\sim t_*$, although $t_{\rm onset}\sim t_{\rm bo}$ when the CSM is denser. The CSM interaction ends when the shock reaches the edge of the CCSM, and its timescale is given by
\begin{eqnarray}
t_{\rm end}\approx
\left\{ \begin{array}{ll} 
1.5\times{10}^{6}~{\rm s}~D_{*,-2}^{1/9}M_{\rm ej,1}^{7/18}{\mathcal E}_{\rm ej,51}^{-1/2}R_{\rm cs,15}^{10/9}.
& \mbox{($\delta=12$)}\\
3.4\times{10}^{7}~{\rm s}~D_{*}^{1/7}M_{\rm ej,1}^{5/14}{\mathcal E}_{\rm ej,51}^{-1/2}R_{\rm cs,16}^{8/7}.
& \mbox{($\delta=10$)}
\end{array} \right.. \,\,\,
\end{eqnarray}

The CR acceleration time is given by $t_{\rm acc}\approx\eta \varepsilon_p/(eBc)$, where $\eta=(20/3)(c^2/V_s^2)$ for a nonrelativistic shock whose normal is parallel to the magnetic field in the Bohm limit~\cite{Drury:1983zz}. The magnetic field is parametrized as
\begin{equation}
U_{B}=\frac{B^2}{8\pi}=\varepsilon_B\frac{3L_s}{4\pi f_\Omega R_s^2 V_s} 
\end{equation}
where $U_B$ is the magnetic energy density. Observations of SNe and numerical simulations suggest $\varepsilon_B\sim{10}^{-3}-{10}^{-2}$~\cite{Maeda:2012pv,Caprioli:2014tva}, and we adopt $\varepsilon_B={10}^{-2}$ as a fiducial value. The magnetic field strength is estimated to be 
\begin{eqnarray}
B\simeq
\left\{ \begin{array}{ll} 
39~{\rm G}~\varepsilon_{B,-2}^{1/2}D_{*,-2}^{1/2}t_{5.5}^{-1}.
& \mbox{($\delta=12$)}\\
380~{\rm G}~\varepsilon_{B,-2}^{1/2}D_{*}^{1/2}t_{5.5}^{-1}.
& \mbox{($\delta=10$)}
\end{array} \right..
\end{eqnarray}

CR acceleration is limited by the age or particle escape if energy losses are irrelevant. In the escape-limited case~\cite{Ohira:2009rd}, for a CR proton with energy $\varepsilon_p$, the maximum energy is
\begin{eqnarray}
\varepsilon_p^{\rm max-esc}\simeq
\left\{\begin{array}{ll} 
9.7 \times{10}^{6}~{\rm GeV}~{(l_{\rm esc}/R_s)}\varepsilon_{B,-2}^{1/2}D_{*,-2}^{3/10}\\
\times{\mathcal E}_{\rm ej,51}^{9/10}M_{\rm ej}^{-7/10}{t}_{5.5}^{-1/5} 
\,\,\, \,\,\, \,\,\, 
\mbox{(for $\delta=12$)}  \,\,\,    \\
4.4 \times{10}^{7}~{\rm GeV}~{(l_{\rm esc}/R_s)}\varepsilon_{B,-2}^{1/2}D_{*}^{1/4}\\
\times{\mathcal E}_{\rm ej,51}^{7/8}M_{\rm ej}^{-5/8}{t}_{5.5}^{-1/4} 
\,\,\, \,\,\, \,\,\, \,\,\, 
\mbox{(for $\delta=10$)}  \,\,\,  
\end{array} \right.,  \,\,\,
\end{eqnarray}
where $l_{\rm esc}$ is the upstream escaping boundary that can be determined by plasma processes including the magnetic field amplification by (nonresonant) CR streaming instabilities~\cite{Inoue:2021bjx,Cristofari:2022low,Diesing:2023ldd} and neutral-ion damping~\cite{Murase:2010cu}. 
In this work, for simplicity, we assume that the escape boundary is comparable to the system size, which is sufficient for the purpose of this work because electromagnetic cascades make the results on photon spectra insensitive to the CR maximum energy. 
In general, other energy losses such as the photomeson production and the Bethe-Heitler pair production process ($p\gamma\rightarrow e^-e^+$) can play roles. In our setup, the inelastic $pp$ interaction is the most relevant cooling process, and we obtain
\begin{eqnarray}
\varepsilon_p^{{\rm max}-{pp}}\simeq
\left\{ \begin{array}{ll} 
1.2\times{10}^{7}~{\rm GeV}~\varepsilon_{B,-2}^{1/2}D_{*,-2}^{-9/10}\\
\times{\mathcal E}_{\rm ej,51}^{9/5}M_{\rm ej}^{-7/5}{t}_{5.5}^{3/5}
\,\,\,
\mbox{(for $\delta=12$)} \,\,\,  \\
2.5\times{10}^{5}~{\rm GeV}~\varepsilon_{B,-2}^{1/2}D_{*}^{-1}\\
\times{\mathcal E}_{\rm ej,51}^{7/4}M_{\rm ej}^{-5/4}{t}_{5.5}^{1/2}
\,\,\,
\mbox{(for $\delta=10$)}  \,\,\,
\end{array} \right.. 
\end{eqnarray}

For CR injections, we assume a CR spectrum to be a power law, i.e., 
\begin{equation}
\frac{dn_{\rm cr}}{dp_{\rm cr}} \propto p_{\rm cr}^{-s_{\rm cr}}e^{-p_{\rm cr}/p_{\rm cr}^{\rm max}},
\end{equation}
where $s_{\rm cr}$ is the CR spectral index and $p_{\rm cr}$ is the CR momentum. We assume proton CRs, and the spectrum is normalized via the CR energy density $U_{\rm cr}$ as
\begin{equation}
U_{\rm cr}=\int d p_p \varepsilon_p \frac{dn_{\rm cr}}{dp_p} =\epsilon_p \frac{1}{2}\varrho_{\rm cs}V_s^2,
\end{equation}
where $\epsilon_{\rm cr}$ is the energy fraction carried by CRs and $\epsilon_{\rm cr}\sim0.1$ is consistent with observations of Galactic CRs~\cite{Murase:2018utn} and fast nova shocks from RS~Ophiuchi~\cite{HESS:2022qap}. 
Although $\epsilon_{\rm cr}$ may be as low as $\epsilon_{\rm cr}\sim0.01$ for radiative shocks, as inferred from observations of classical novae~\cite{Li:2017crr} and constraints from SN 2010jl~\cite{Murase:2010cu}, the shock is adiabatic for modest values of $D_*$ including the cases without CCSM, and the CR acceleration could also be modified by the preacceleration of CRs at the CSM eruption~\cite{Murase:2013kda}. We use $\epsilon_{\rm cr}=0.1$, which is sufficient for the purpose of this work. 
Although we assume $s_{\rm cr}=2.0-2.2$, we present the results for $s_{\rm cr}=2.2$ throughout this work because Galactic CR data prefer $s_{\rm cr}\sim2.2-2.4$ rather than $s_{\rm cr}=2.0$~\cite{Murase:2018utn}. 
However, we note that our results are largely insensitive to $s_{\rm cr}$ thanks to electromagnetic cascades that make photon energy spectra ``flat''. See also M18 for the impacts on the detectability of high-energy neutrinos. 

CR ions interact with cold nucleons in the CSM and lead to the $pp$ production of mesons (mostly pions), which generates a flux of high-energy neutrinos via decay processes like $\pi^+\to\mu^+\nu_\mu\to\nu_\mu\bar{\nu}_\mu\nu_ee^+$ and gamma rays via $\pi^0\rightarrow2\gamma$. The typical neutrino energy is $\varepsilon_\nu\sim(0.03-0.05)\varepsilon_p$~\cite{Kelner:2006tc}. The approximate cross section and proton inelasticity of $pp$ collisions are $\sigma_{pp}\approx3\times{10}^{-26}~{\rm cm}^2$ and $\kappa_{pp}\approx0.5$, respectively. Using Eqs.~(\ref{RsIIP}) and (\ref{VsIIP}), the effective optical depth to inelastic $pp$ interactions, $f_{pp}\approx\kappa_{pp}\sigma_{pp}(\varrho_{\rm cs}/m_H)R_{s}(c/V_s)$, is estimated to be~\cite{Murase:2017pfe}
\begin{eqnarray}
f_{pp} \simeq
\left\{\begin{array}{ll} 
0.82~D_{*,-2}^{6/5}M_{\rm ej,1}^{7/10}{\mathcal E}_{\rm ej,51}^{-9/10}t_{5.5}^{-4/5}.
& \mbox{($\delta=12$)}\\
170~D_{*}^{5/4}M_{\rm ej,1}^{5/8}{\mathcal E}_{\rm ej,51}^{-7/8}t_{5.5}^{-3/4}.
& \mbox{($\delta=10$)}
\end{array} \right.
\label{eq:fpp}
\end{eqnarray}
This gives the transition time, at which the system becomes effectively optically thin for confined CR ions, 
\begin{eqnarray}
t_{f_{pp}=1}\simeq
\left\{ \begin{array}{ll} 
2.4\times10^{5}~{\rm s}~D_{*,-2}^{3/2}M_{\rm ej,1}^{7/8}{\mathcal E}_{\rm ej,51}^{-9/8}
& \mbox{($\delta=12$)}\\
3.1\times10^{8}~{\rm s}~D_{*}^{5/3}M_{\rm ej,1}^{5/6}{\mathcal E}_{\rm ej,51}^{-7/6}
& \mbox{($\delta=10$)}.
\end{array} \right.
\label{eq:tfpp}
\end{eqnarray}
One sees that the dense CSM allows us to naturally expect that early SNe have the bright phase in high-energy neutrinos and gamma rays.

\subsection{Thermal emission}
%
\begin{figure}[t]
\includegraphics[width=\linewidth]{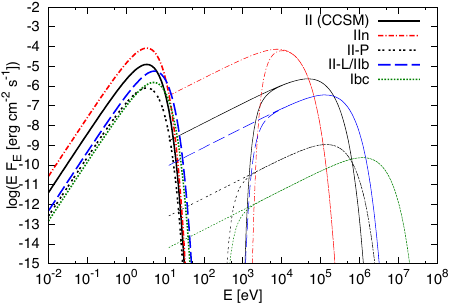}
\caption{Energy fluxes of thermal optical (thick) and x-ray (thin) emission used in this work. For thermal bremsstrahlung emission, spectra without/with external matter attenuation are shown with upper/lower curves. A Galactic SN at $d=10$~kpc is considered, and $D_*=0.01$ is used for a SN II (CCSM). 
\label{fig:thspectrum}
}
\end{figure}

{\sc AMES} focuses on the calculations of nonthermal emission but thermal emission has to be modeled consistently especially for the studies of SNe. This is because thermal photons typically overwhelm nonthermal photons in the optical and x-ray bands, and they provide target photons for high-energy gamma rays through $\gamma\gamma\rightarrow e^+e^-$.  
We approximately implement time-dependent thermal spectra in the following physical manner. 

The kinetic energy of SN ejecta is dissipated through shocks. The forward shock is more important for our ejecta profiles~\cite{Chevalier:2016hzo} and CSM parameters ($M_{\rm cs}<M_{\rm ej}$) considered in this work (but see, e.g., Ref.~\cite{Murase:2010cu} for reverse shock contributions). Then, the postshock temperature ${\mathcal T}'_{\rm cs}$ is
\begin{eqnarray}
k{\mathcal T}'_{\rm cs}&=&\frac{3}{16}m_H\mu V_s^2\\
&\simeq&
\left\{ \begin{array}{ll} 
57~{\rm keV}~D_{*,-2}^{-1/5}M_{\rm ej,1}^{-7/10}{\mathcal E}_{\rm ej,51}^{9/10}t_{5.5}^{-1/5}\\
\mbox{(for $\delta=12$)} \,\,\, \,\,\, \,\,\, \,\,\, \,\,\, \,\, \,\,\, \,\,\, \,\,\, \,\,\, \,\,\, \,\,\, \,\,\, \,\,\,  \,\,\, \\
26~{\rm keV}~D_{*}^{-1/4}M_{\rm ej,1}^{-5/8}{\mathcal E}_{\rm ej,51}^{7/8}t_{5.5}^{-1/4}\\
\mbox{(for $\delta=10$)}  \,\,\, \,\,\, \,\,\, \,\,\, \,\,\, \,\, \,\,\, \,\,\, \,\,\, \,\,\, \,\,\, \,\,\, \,\,\, \,\,\,  \,\,\, 
\end{array} \right.,
\end{eqnarray}
which is typically in the hard x-ray range, and $\mu$ is the mean molecular weight and $m_H$ is the hydrogen mass. 
Note that the immediate downstream temperature can be reduced if the Compton cooling is faster than plasma heating processes such as the Coulomb heating from ions to electrons~\cite{Murase:2010cu,Katz:2012zzk, Chevalier:2012zf}. 
The thermal bremsstrahlung luminosity is 
\begin{eqnarray}
L_{\rm bre}&=&4 \pi R^2 \Delta R' (\Lambda_{\rm ff} n'_e n'_H)\nonumber\\
&\simeq&2.9\times{10}^{41}~{\rm erg}~{\rm s}^{-1}~\mu_H^{-1}\mu_e^{-1}D_{*,-2}^{2}t_{5.5}^{-1},
\end{eqnarray}
where $\Lambda_{\rm ff}$ is the cooling function. For moderate values of $D_*\lesssim0.03$, including SN 2013fs-like cases, the shock is expected to be adiabatic. If $L_{\rm bre}>L_{s}$, where the shock would be radiative, we limit the bremsstrahlung luminosity by $L_s$. 

In dense environments, photons are reprocessed both in the downstream and upstream, and a significant energy fraction of the radiation should be released as SN emission in the optical band. In this work, following M18, we implement a gray body spectrum with thermal luminosity $L_{\rm sn}=\epsilon_{\rm rad}L_s$ for a component originating from the interaction with CSM. The photon temperature is set by ${\mathcal T}_{\rm sn}={\rm max}[{\mathcal T}_{\rm bb},{\mathcal T}_{\rm rec}]$, where ${\mathcal T}_{\rm bb}$ is the black body temperature and ${\mathcal T}_{\rm rec}={10}^4$~K, and $\epsilon_{\rm rad}=1/4$ is used. In addition, there can be an ordinary SN component from the photosphere, which can be energized by radioactive nuclei, shocks propagating in the progenitor, and resulting cooling envelope emission. We also add this external component with luminosity $L_{\rm ph}$ to discuss the impacts. As light-curve templates, optical luminosities of SN 1999bm for Type II-P~\cite{BH09} and SN 2004aw for Type Ibc~\cite{Mazzali:2017eqq} are considered in our baseline calculations.

\subsection{Numerical method}
For given dynamics and CR acceleration, {\sc AMES} allows us to numerically calculate neutrino and gamma-ray spectra through solving kinetic equations in a time-dependent manner as in M18. For SN dynamics, we employ the equation of motion of the shocked shell for parameters listed in Table~\ref{tab:CCSM}, and we evaluate $\varepsilon_p^{\rm max}$. The SN module of {\sc AMES} can be used for arbitrary $R_s$, $V_s$, $\varrho_{\rm cs}$, $L_{\rm sn}$, CR distributions, and the magnetic field as a function of time. Although {\sc AMES} includes various processes for particle production, $pp$ interactions, $p\gamma$ interactions, nuclear photodisintegration, and Bethe-Heitler pair production, in this work it is sufficient to consider $pp$ interactions. See Ref.~\cite{Zhang:2023ewt} for details on the other processes. Compared to previous works, in {\sc AMES} we have updated treatments on $pp$ interactions, Coulomb losses, bremsstrahlung, and free-free absorption. 

Distributions of gamma rays and electrons/positrons are obtained by solving the following partial differential equations (see Supplementary Material of M18 for details) via the implicit method, 
\begin{eqnarray}\label{eq:cascade}
\dot{n}_{\varepsilon_e}^e &=& \dot{n}_{\varepsilon_e}^{(\gamma\gamma)}
- \frac{\pd}{\pd \varepsilon_e} [(P_{\rm IC}+P_{\rm syn}+P_{\rm ad}+P_{\rm bre}+P_{\rm Cou}) n_{\varepsilon_e}^e] \nonumber\\ 
& & + \dot{n}_{\varepsilon_e}^{\rm inj}\nonumber\\ 
\dot{n}_{\varepsilon_\gamma}^\gamma &=& -\frac{n_{\varepsilon_\gamma}^{\gamma}}{t_{\gamma \gamma}} 
- \frac{n_{\varepsilon_\gamma}^{\gamma}}{t_{\rm esc}} -\frac{n_{\varepsilon_\gamma}^{\gamma}}{t_{\rm matter}} 
+ \dot{n}_{\varepsilon_\gamma}^{(\rm IC)}
+ \dot{n}_{\varepsilon_\gamma}^{(\rm syn)}
+ \dot{n}_{\varepsilon_\gamma}^{(\rm bre)} \nonumber\\ 
& & + \dot{n}_{\varepsilon_\gamma}^{\rm inj}\nonumber\\ 
\end{eqnarray}
where $n_{\varepsilon_i}^i\equiv dn^i/d\varepsilon_i$ and $\varepsilon_i$ is particle energy for a particle species with $i$. Energy loss rates of electrons/positrons for the IC radiation ($P_{\rm IC}$), synchrotron radiation ($P_{\rm syn}$), adiabatic cooling ($P_{\rm ad}$), relativistic bremsstrahlung ($P_{\rm bre}$), and Coulomb collisions ($P_{\rm Cou}$), respectively, are~\cite{Blumenthal:1970gc,Schlickeiser:2002pg}
\begin{eqnarray}
P_{\rm IC}&=& \int d \varepsilon_\gamma \,\, \varepsilon_\gamma 
\int d \varepsilon_t \,\, n_{\varepsilon_t}^\gamma \int \frac{d \cos\theta}{2} \,\, \tilde{c} \frac{d \sigma_{\rm IC}}{d \varepsilon_\gamma}
, \nonumber \\
P_{\rm syn}&=&\frac{1}{6\pi}\sigma_T c \frac{p_e^2}{m_e^2c^2}B^2
, \nonumber \\
P_{\rm bre}&=&\left(\sum_i \frac{Z_i^2}{A_i}x_i+\frac{1}{\mu_e}\right)\frac{3\alpha_{\rm em} }{2\pi} \sigma_T c n'_N \left(\ln[2\gamma_e]-\frac{1}{3}\right)\varepsilon_e
, \nonumber \\
P_{\rm Cou}&=&\frac{3}{4}\sigma_T m_e c^3 n'_{N} \mu_e^{-1} \left(74.3 + \ln[\gamma_e\mu_e/n'_{N}] \right),
\end{eqnarray}
where $\tilde{c} =(1-\cos\theta)c$ (where $\theta$ is the angle between two particles), $d\sigma_{\rm IC}/d\varepsilon_\gamma$ is the differential IC cross section for target photon energy $\varepsilon_t$~\cite{Blumenthal:1970gc}, $Z_i$ is the nuclear charge, $A_i$ is the nuclear mass number, $x_i$ is the nuclear mass fraction, and $\mu_e^{-1}$ is the number of electrons per baryon. The weak shielding limit is assumed for bremsstrahlung and collisions with electrons in the shell are considered, and contributions from both nuclei and electrons are included~\cite{Schlickeiser:2002pg}.  
High-energy gamma rays interact with other photons via $\gamma \gamma \rightarrow e^+e^-$, and we take into account the gamma-ray attenuation and subsequent regeneration. The two-photon annihilation rate and cross section are
\begin{eqnarray}
t_{\gamma \gamma}^{-1} &=& \int d \varepsilon_t \,\, n_{\varepsilon_t}^\gamma  \int \frac{d \cos\theta}{2} \,\, \tilde{c} \sigma_{\gamma \gamma} 
, \nonumber \\
\sigma_{\gamma \gamma}&=&\frac{3}{16} \sigma_T (1-\beta_{\rm cm}^2) \left[2 \beta_{\rm cm} (\beta_{\rm cm}^2-2) \right.\nonumber\\
&+& \left.(3-\beta_{\rm cm}^4) \ln[(1+\beta_{\rm cm})/(1-\beta_{\rm cm})] \right],
\end{eqnarray}
respectively, where $\beta_{\rm cm}=\sqrt{(1-4 m_e^2 c^4/S)}$, $S$ is the Mandelstam variable, and $t_{\rm esc}=R_s/c$ is the photon escape time. Differential particle generation rate densities are
\begin{eqnarray}
\dot{n}_{\varepsilon_e}^{(\gamma \gamma)} &=& \frac{1}{2} \int d \varepsilon_\gamma\,\, n_{\varepsilon_\gamma}^\gamma \, \int d \varepsilon_t\,\, n_{\varepsilon_t}^\gamma \, \int \frac{d \cos\theta}{2} \,\, \tilde{c}  \frac{d \sigma_{\gamma \gamma}}{d \varepsilon_e} 
, \nonumber \\
\dot{n}_{\varepsilon_\gamma}^{(\rm IC)} &=& \int d \varepsilon_e \,\, n_{\varepsilon_e}^e \, \int d \varepsilon_t\,\, n_{\varepsilon_t}^\gamma \, \int  \frac{d \cos\theta}{2} \,\, \tilde{c}   \frac{d \sigma_{\rm IC}}{d \varepsilon_\gamma} 
, \nonumber \\
\dot{n}_{\varepsilon_\gamma}^{(\rm syn)} &=& \int d \varepsilon_e \,\, n_{\varepsilon_e}^e \, \frac{1}{\varepsilon_\gamma} \frac{d P_{\rm syn}}{d \varepsilon_\gamma} 
, \nonumber \\
\dot{n}_{\varepsilon_\gamma}^{(\rm bre)} &=& \int d \varepsilon_e \,\, n_{\varepsilon_e}^e \, \frac{1}{\varepsilon_\gamma} \frac{d P_{\rm bre}}{d \varepsilon_\gamma} 
,
\end{eqnarray}
where $d\sigma_{\gamma\gamma}/d\varepsilon_e=2\sigma_{\gamma\gamma}\delta(\varepsilon_e-\varepsilon_\gamma/2)$ is the simplified differential two-photon annihilation cross section with $\varepsilon_e=\varepsilon_\gamma/2$ as used as in Ref.~\cite{Murase:2014bfa}, and $dP_{\rm syn}/d\varepsilon_\gamma$ and $dP_{\rm bre}/d\varepsilon_\gamma$ are synchrotron and bremsstrahlung~\citep{Blumenthal:1970gc,Zhang:2020qbt} powers per photon energy, respectively. 

Photons lose their energies due to interactions with matter during their escape from the shocked shell, and $t_{\rm matter}^{-1}=(\kappa_{\rm BH} \sigma_{\rm BH}+\kappa_{\rm Comp} \sigma_{\rm Comp})n_Nc$ is the energy loss time and $n_N$ is the average nucleon density. The attenuation cross section of the Bethe-Heitler pair production process can be approximated by
\begin{equation}
\kappa_{\rm BH} \sigma_{\rm BH}=\frac{x-2}{x}\sigma_{\rm BH},
\end{equation}
where $x\equiv \varepsilon_\gamma/(m_ec^2)$. Ignoring the contribution from electron-positron annihilation, we use inelasticity $\kappa_{\rm BH}$ and cross section $\sigma_{\rm BH}$ obtained with the Born approximation~\cite{CZS92}.  
The attenuation cross section of the Compton scattering process is~\cite{Murase:2014bfa}
\begin{eqnarray}
\kappa_{\rm Comp} \sigma_{\rm Comp}&=&\frac{3}{4}\sigma_T\left[\frac{2{(1+x)}^2}{x^2(1+2x)}-\frac{1+3x}{{(1+2x)}^2}\right.\nonumber\\
&+&\frac{(1+x)(2x^2-2x-1)}{x^2{(1+2x)}^2}+\frac{4x^2}{3{(1+2x)}^3}\nonumber\\
&+&\left.\left(\frac{1+x}{x^3}-\frac{1}{2x}+\frac{1}{2x^3}\right)\ln(1+2x) \right],
\end{eqnarray}
where the Klein-Nishina effect is fully taken into account. 

The differential injection rate densities via $pp$ interactions are calculated by 
\begin{eqnarray}
\dot{n}_{\varepsilon_\nu}^{\rm inj}=\frac{d\sigma_{pp}\xi_\nu}{d\varepsilon_\nu} \frac{cM_{\rm cs}(<R_s)}{m_H\mathcal V} \int dp_{\rm cr} \,\,\, \frac{dn_{{\rm cr}}}{dp_{\rm cr}}\nonumber\\
\dot{n}_{\varepsilon_\gamma}^{\rm inj}=\frac{d\sigma_{pp}\xi_\gamma}{d\varepsilon_\gamma} \frac{cM_{\rm cs}(<R_s)}{m_H\mathcal V}\int dp_{\rm cr} \,\,\, \frac{dn_{{\rm cr}}}{dp_{\rm cr}}\nonumber\\
\dot{n}_{\varepsilon_e}^{\rm inj}=\frac{d\sigma_{pp}\xi_e}{d\varepsilon_e} \frac{cM_{\rm cs}(<R_s)}{m_H\mathcal V} \int dp_{\rm cr} \,\,\, \frac{dn_{{\rm cr}}}{dp_{\rm cr}},
\end{eqnarray}
where $\xi_\nu$, $\xi_\gamma$ and $\xi_e$ are multiplicities of neutrinos, gamma rays and electrons/positrons, respectively, $\mathcal V$ is the volume, and $M_{\rm cs}(<R_s)$ is the CSM mass swept up by the forward shock. In {\sc AMES}, treatments on low-energy neutrinos and gamma rays are improved by interpolating simulation results of {\sc Geant4}~\cite{GEANT4:2002zbu} with those from differential spectra of secondary particles following the parametrization by Ref.~\cite{Kelner:2006tc} and the $pp$ cross section by Ref.~\cite{Kafexhiu:2014cua}. The resulting secondary spectra are consistent with the previous works~\cite{Kamae:2006bf,Kafexhiu:2014cua}.

Intrinsic energy fluxes of neutrino and gamma rays reaching Earth are given by 
\begin{eqnarray}
E F_{E}=\frac{\varepsilon L_{\varepsilon}}{4\pi d_L^2}=\frac{(\varepsilon^2 n_{\varepsilon})\mathcal V}{t_{\rm esc}},
\end{eqnarray}
where $E=\varepsilon/(1+z)$ is particle energy on Earth, and $d_L$ is the luminosity distance. 
For neutrinos, the flavor mixing is additionally taken into account. 
For photons, we implement the synchrotron self-absorption (SSA) process by multiplying $\tau_{\rm sa}^{-1}(1-e^{-\tau_{\rm sa}})$, where   
\begin{equation}\label{eq:taugammagamma}
\tau_{\rm sa}=R_{s}\int d \gamma_e \, n_{\gamma_e}^e\sigma_{\rm sa},
\end{equation}
is the SSA optical depth in the emission region, and the SSA cross section is~\citep[e.g.,][]{Chiaberge:1998cv,GS91}.
\begin{equation}
\sigma_{\rm sa}=\frac{1}{8\pi m_e\nu^2\gamma_ep_e}\frac{\partial}{\partial \gamma_e}\left[\gamma_ep_e\frac{dP_{\rm syn}}{d\nu}\right],
\end{equation}
where $\nu=\varepsilon_\gamma/h$ is the frequency. 

Photons leaving the source are further reprocessed in the external screen region, which corresponds to unshocked CSM in our setup. 
We treat such external attenuation by the suppression factor, $f_{\rm sup}^{\gamma\gamma-\rm ex}=e^{-\tau_{\gamma \gamma}}$. This is reasonable when photons are significantly reprocessed to optical emission in the CSM. For the photons coming from the photosphere this treatment is more approximate. For matter attenuation of photons, as in Ref.~\cite{Murase:2018okz}, we implement 
\begin{equation}
f_{\rm sup}^{\rm mat-ex}={\rm max}[e^{-\tau_{\rm mat}},f_{\rm esc}].
\end{equation}
where $\tau_{\rm mat}$ consists of all processes, Bethe-Heitler pair production, Compton, photoelectric absorption and free-free absorption. We also introduce $f_{\rm esc}$ as the effective escape fraction, which can be significant when the CSM is clumpy or aspherical.  

At sufficiently high energies above the electron-positron pair production threshold, the Bethe-Heitler pair production process is dominant. The Bethe-Heitler cross section on a nucleus scales as $\sigma_{\rm BH}=Z^2\sigma_{\rm BH}^{(p)}$, where $\sigma_{\rm BH}^{(p)}$ is the cross section on a proton. Taking into account contributions from both nuclei and electrons, we have
\begin{eqnarray}
\kappa_{\rm BH}\tau_{\rm BH}&=&\left(\sum_i \frac{Z_i^2}{A_i}x_i+\frac{1}{\mu_e}\right)\frac{D \kappa_{\rm BH}\sigma_{\rm BH}^{(p)}}{m_HR_s}\nonumber\\
&=&(\tilde{Z}+1)\kappa_{\rm BH}\sigma_{\rm BH}^{(p)}n_e R_s,  
\end{eqnarray}
where $\tilde{Z}\equiv\Sigma_i(Z_i^2/A_i)x_i\mu_e$, which depends on chemical composition of the ejecta. We consider $x_{\rm H}=0.7$ and $x_{\rm He}=0.25$ and $x_{\rm CNO}=0.05$, leading to $\tilde{Z}\simeq0.97$ and $\mu_e\simeq0.85$.  

Energy losses of x-ray and gamma-ray photons via Compton scattering are included as effective attenuation considering the inelasticity. The effective optical depth is given by
\begin{equation}
\kappa_{\rm Comp}\tau_{\rm Comp}=\frac{D \kappa_{\rm Comp}\sigma_{\rm Comp}}{\mu_e m_HR_s},
\end{equation}
where $\kappa_{\rm Comp}\sigma_{\rm Comp}$.
In addition, in the x-ray band, we implement photoelectric absorption with its optical depth, $\tau_{\rm pe}=K_{\rm pe}\varrho_{\rm cs}R_s$, where we use $K_{\rm pe}=0.024~{\rm g}^{-1}~{\rm cm}^{2}~{(\varepsilon/1~\rm keV)}^{-3}$ above 10.2~keV for simplicity. 
Nonthermal emission can overwhelm thermal emission at sufficient high energies, where the photoelectric absorption is not very important.    

Free-free absorption can be crucial at the radio band. The free-free optical depth is~\citep{MH67} 
\begin{eqnarray}
\tau_{\rm ff}=\int_{ R_s} dR \, \alpha_{\rm ff}&\approx&\frac{1}{(2w-1)}\sum_i 8.4\times{10}^{-28}~\frac{\hat{Z}_i^2}{A_i}x_i \nonumber\\
&\times& {\mathcal T}_{\rm cs,4}^{-1.35}\nu_{10}^{-2.1}n_e n_N R_s,\,\,\,\,\,\,\,\,\,\,\,\,
\end{eqnarray}
where ${\mathcal T}_{\rm cs}$ is the upstream CSM temperature and $\hat{Z}$ is the effective charge considering the ionization. Radio waves may be absorbed by cold plasma in the upstream and we also include the Razin effect in an approximate manner with an exponential cutoff, where the Razin-Tsytovich frequency is given by $\nu_{\rm RT}\equiv 2e c n_e/B$.   

\begin{figure*}[tb]
\includegraphics[width=0.49\linewidth]{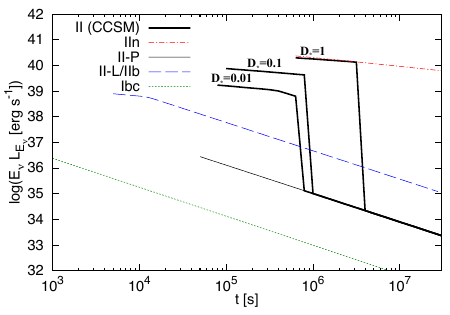}
\includegraphics[width=0.49\linewidth]{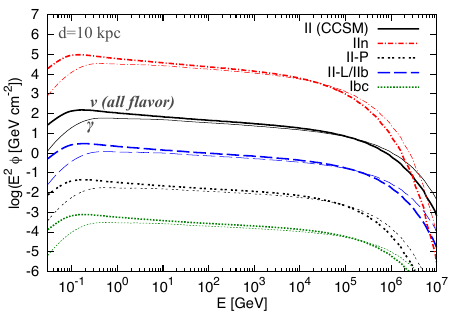}
\caption{Left panel: light curves of high-energy neutrinos (at $\varepsilon_\nu=1$~TeV) for various types of SNe. Right panel: energy fluences of $\nu_e+\bar{\nu}_e+\nu_\mu+\bar{\nu}_\mu+\nu_\tau+\bar{\nu}_\tau$ (thick) and generated gamma rays (thin) integrated over $t_{\nu \rm max}$. A Galactic SN at $d=10$~kpc is considered and $D_*=0.01$ is used for a SN II (CCSM). In both panels, $s_{\rm cr}=2.2$ is assumed.  
\label{fig:nuLCsp}
}
\end{figure*}

\section{Light curves and spectra}\label{sec:lcspe}
We calculate multimessenger spectra and light curves that are obtained by numerical calculations with {\sc AMES}. We also present analytical expressions of multimessenger spectra to understand the physical situation, assuming $\delta=12$. Throughout this work, the total SN kinetic energy (i.e., integrated over velocities) and SN ejecta mass are set to ${\mathcal E}_{\rm ej}=10^{51}~{\rm erg}~{\mathcal E}_{\rm ej,51}$ and $M_{\rm ej}=10~M_{\odot}~{M}_{1}$, respectively. 

\subsection{Neutrinos}
The differential neutrino luminosity (for the sum of all flavors) above $\sim1$~GeV is approximated to be
\begin{eqnarray}
\varepsilon_{\nu} L_{\varepsilon_{\nu}}&\approx&\frac{1}{2}{\rm min}[f_{pp},1] \frac{\epsilon_{\rm cr} L_s}{{\mathcal R}_{\rm cr10}}{\left(\frac{\varepsilon_\nu}{0.4~{\rm GeV}}\right)}^{2-s_\nu}\nonumber\\
&\simeq&5.0\times{10}^{39}~{\rm erg}~{\rm s}^{-1}~{\rm min}[f_{pp},1]f_\Omega\nonumber\\
&\times&\epsilon_{\rm cr,-1}{\mathcal R}_{\rm cr10,1}^{-1}{\left(\frac{\varepsilon_\nu}{0.4~{\rm GeV}}\right)}^{2-s_\nu}\nonumber\\
&\times&D_{*,-2}^{7/10}{\mathcal E}_{\rm ej,51}^{27/20}M_{\rm ej,1}^{-21/20}t_{5.5}^{-3/10},
\label{eq:nulum}
\end{eqnarray}
where the factor $1/2$ comes from the facts that the $\pi^\pm/\pi^0$ ratio is $\approx2$ in $pp$ interactions and neutrinos carry $3/4$ of the pion energy in the decay chain. We also introduce ${\mathcal R}_{\rm cr10}\equiv\epsilon_{\rm cr}L_s/(\varepsilon_p^2 d\dot{N}_{\rm cr}/d\varepsilon_p)|_{10~{\rm GeV}}$ that is a spectrum dependent factor that converts the bolometric luminosity to the differential luminosity, and $s\approx s_{\rm cr}$ for $f_{pp}\gtrsim1$ and $s\approx s_{\rm cr}-0.1$ for $f_{pp}\lesssim1$. In the latter case, the secondary spectra are somewhat harder than the CR spectrum because of the weak energy dependence of $pp$ interactions. 

Given that $pp$ interactions are dominant, the neutrino flux would obey the following scaling,
\begin{equation}
E_\nu F_{E_\nu}
\propto
\left\{
\begin{array}{lr} 
\varepsilon_\nu^{2-s_{\rm cr}}t^{-0.3} & \mbox{($t \leq t_{f_{pp}=1}$)} \\
\varepsilon_\nu^{2.1-s_{\rm cr}}t^{-1.1} & \mbox{($t > t_{f_{pp}=1}$)} 
\end{array} 
\right., 
\label{eq:nuscaling}
\end{equation}
where $f_{pp}$ is given by Eq.~(\ref{eq:fpp}). For neutrino detection, the time evolution of the energy fluence is important as long as the atmospheric background is negligible~\cite{Murase:2017pfe}. Equation~(\ref{eq:nuscaling}) suggests that the neutrino fluence has a peak around $t_{f_{pp}=1}$, which gives a characteristic timescale of high-energy neutrino emission. 

In Fig.~\ref{fig:nuLCsp} left, we show neutrino light curves at $\varepsilon_\nu=1$~TeV. Thick curves represent CCSN-interacting SNe, while thin curves do SNe II-P without CCSM. We use $R_{\rm cs}=4\times{10}^{14}$~cm for $D_*=0.01$ and $D_*=0.1$ as in M18, while we consider $R_{\rm cs}={10}^{15}$~cm and $D_*=1$ to cover the most optimistic cases (see also Ref.~\cite{Kheirandish:2022eox}). Our model predicts that the neutrino luminosity typically ranges from $L_{\nu}\sim10^{39}-{10}^{40}~{\rm erg}~{\rm s}^{-1}$ with durations of $\sim1-30$~d. With CCSM, the system is nearly calorimetric so that the neutrino luminosity scales as $D^{7/10}$. The temporal slope changes at the time when $f_{pp}$ becomes less than unity, which is consistent with Eq.~(\ref{eq:nuscaling}). The onset time is significantly longer for $D_*=1$ because the shock propagating in the CSM is radiation mediated at early times. For Type IIn SNe, characteristic time windows are $\sim0.1-1$~yr after the explosion~\cite{Murase:2010cu}. 
Without CCSM, $t_{\rm onset}$ almost coincides with shock breakout from the progenitor. We note that as shown in M18 high-energy neutrinos from Galactic SNe are detectable for IceCube, KM3Net~\cite{KM3Net:2016zxf}, Baikal-GVD~\cite{Baikal-GVD:2018isr}, P-ONE~\cite{P-ONE:2020ljt}, TRIDENT~\cite{Ye:2022vbk} and especially IceCube-Gen2~\cite{IceCube-Gen2:2020qha} even without CCSM. 

In Fig.~\ref{fig:nuLCsp} right, we show energy fluences of neutrinos and generated gamma rays at $t_{\nu{\rm max}}$, which is the time when the significance of neutrino emission becomes the maximum. One of the ultimate goals is to reveal CR ion acceleration through multimessenger observations, for which simultaneous observations between neutrinos and photons are critical. We show energy fluxes at $t_{\nu{\rm max}}$ as default, and the values are shown in Table~\ref{tab:CCSM}. 
High-energy SN neutrino emission is long lasting with optimized time windows from days to months for SNe II, which can be regarded as high-energy neutrino afterglows in contrast to SN neutrino bursts in the MeV range.

\subsection{Pionic gamma rays}
%
\begin{figure*}[tb]
\includegraphics[width=0.325\linewidth]{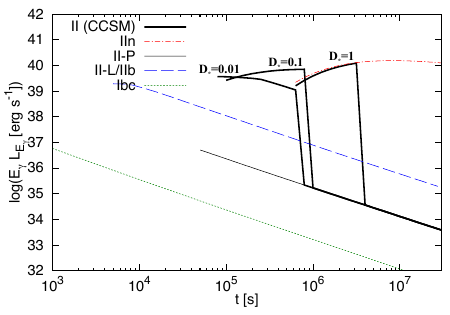}
\includegraphics[width=0.325\linewidth]{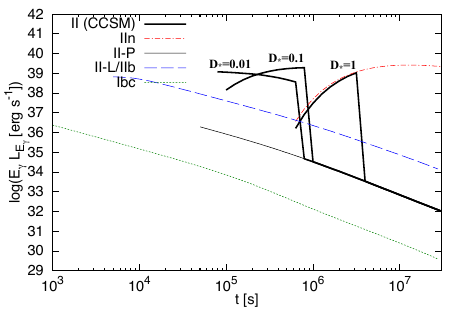}
\includegraphics[width=0.325\linewidth]{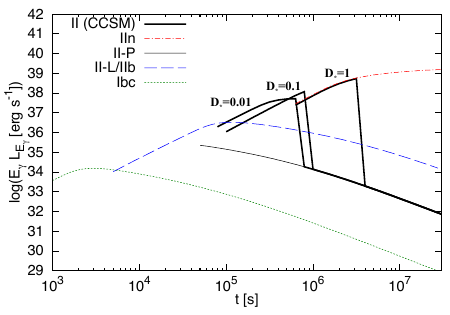}
\caption{Light curves of high-energy gamma rays at $\varepsilon_\gamma=1$~GeV (left panel), MeV gamma rays at $\varepsilon_\gamma=1$~MeV (middle panel), and radio waves at $\nu=100$~GHz (right panel) for various types of SNe. Note that radio emission without external matter attenuation is shown. In all three panels, $s_{\rm cr}=2.2$ is assumed.  
\label{fig:EMLC}
}
\end{figure*}

\begin{figure}[tb]
\includegraphics[width=\linewidth]{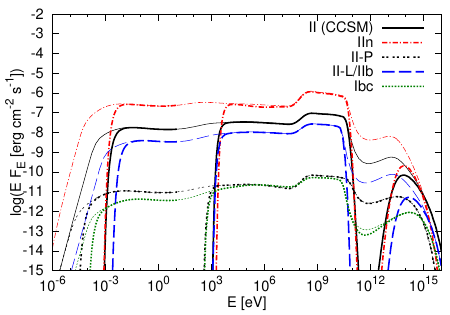}
\caption{Energy fluxes of nonthermal electromagnetic emission from a SN at $d=10$~kpc, corresponding to Fig.~\ref{fig:nuLCsp}. Spectra with/without external matter attenuation (thick/thin) are shown. For a SN II (CCSM), $D_*=0.01$ is used. 
\label{fig:EMsp}
}
\end{figure}

\begin{figure*}[tb]
\includegraphics[width=0.325\linewidth]{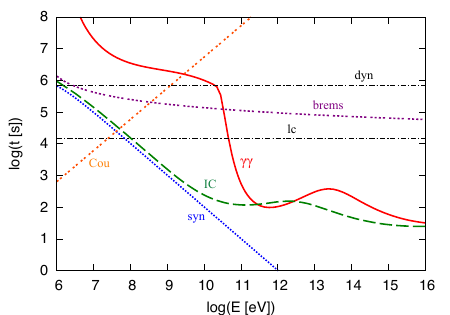}
\includegraphics[width=0.325\linewidth]{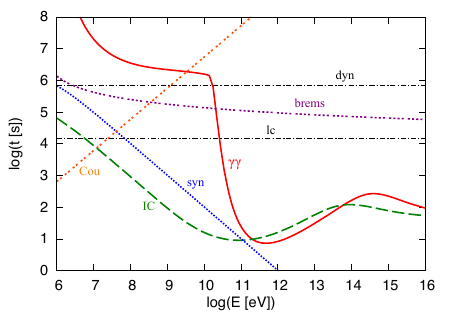}
\includegraphics[width=0.325\linewidth]{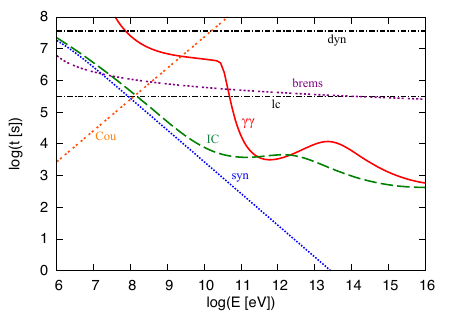}
\caption{Timescales relevant for electromagnetic cascades for CCSM-interacting SNe II without $L_{\rm ph}$ (left panel), CCSM-interacting SNe II with $L_{\rm ph}$ (middle panel) and SNe IIn (right panel). Considered energy loss processes for electrons and positrons are synchrotron radiation (syn), IC scattering (IC), bremmstrahlung emission (brems), Coulomb scattering (Cou) and adiabatic losses whose timescale is comparable to the dynamical time (dyn). For photons, the interaction time of two-photon annihilation ($\gamma\gamma$) and the light crossing time (lc) are shown.   
\label{fig:timescale}
}
\end{figure*}

Energy fluxes of generated gamma rays and neutrinos are related by the multimessenger connection. The differential luminosities of neutrinos and pionic gamma rays (from $\pi^0\rightarrow2\gamma$) are approximately related as~\cite{Murase:2013rfa}
\begin{equation}
\varepsilon_\gamma L_{\varepsilon_\gamma}^{\pi^0}\approx \frac{2}{3}{[\varepsilon_\nu L_{\varepsilon_\nu}]}_{\varepsilon_\nu=\varepsilon_\gamma/2} f_{\rm sup},
\label{eq:mm}
\end{equation}
where $f_{\rm sup}=f_{\rm sup}(\varepsilon_\gamma)$ is the energy-dependent suppression fraction, which considers both emission and screen regions. Equation~(\ref{eq:mm}) suggests that the energy fluxes of neutrinos and generated gamma rays are comparable, which is an unavoidable consequence of inelastic $pp$ interactions. The differential gamma-ray luminosity is 
\begin{eqnarray}
\varepsilon_\gamma L_{\varepsilon_\gamma}^{\pi^0}&\approx&\frac{f_{\rm sup}}{3}{\rm min}[f_{pp},1]\frac{\epsilon_{\rm cr}L_s}{{\mathcal R}_{\rm cr10}} {\left(\frac{\varepsilon_\gamma}{0.8~{\rm GeV}}\right)}^{2-s}\nonumber\\
&\simeq&3.3\times{10}^{39}~{\rm erg}~{\rm s}^{-1}~{\rm min}[f_{pp},1]f_{\rm sup}f_\Omega\nonumber\\
&\times&\epsilon_{\rm cr,-1}{\mathcal R}_{\rm cr10,1}^{-1}{\left(\frac{\varepsilon_\gamma}{0.8~{\rm GeV}}\right)}^{2-s_\gamma}\nonumber\\
&\times&D_{*,-2}^{7/10}{\mathcal E}_{\rm ej,51}^{27/20}M_{\rm ej,1}^{-21/20}t_{5.5}^{-3/10}.
\label{eq:gammalum}
\end{eqnarray}
which agrees with the numerical curves shown in Fig.~\ref{fig:EMLC} left. Photon spectra resulting from spectra of generated gamma rays (see Fig.~\ref{fig:nuLCsp} right) are shown at $t_{\nu{\rm max}}$ in Fig.~\ref{fig:EMsp}, where electromagnetic cascades are included.  

The suppression factor for gamma rays mainly consists of two parts, $f_{\rm sup}^{\rm BH-ex}$ and $f_{\rm sup}^{\gamma\gamma-\rm ex}$. For analytical estimates on the Bethe-Heitler suppression, one may use a simpler formula,
\begin{equation}
\sigma_{\rm BH}\approx Z^2\frac{3\alpha_{\rm em}}{8\pi}\sigma_T\left[\frac{28}{9}\ln(2x)-\frac{218}{27}\right],
\end{equation}
where $\alpha_{\rm em}$ is the fine structure constant, and we see $\sigma_{\rm BH}\sim Z^2{10}^{-26}~{\rm cm}^2$ at GeV energies. 
Then, the Bethe-Heitler optical depth for gamma rays with $\varepsilon_\gamma=1$~GeV is estimated to be
\begin{eqnarray}
\tau_{\rm BH}&\approx&0.031~[(\tilde{Z}+1)/2]\tau_T\nonumber\\
&\simeq&0.025~(\tilde{Z}+1)\mu_e^{-1}D_{*,-2}^{11/10}{\mathcal E}_{\rm ej,51}^{-9/20}M_{\rm ej,1}^{7/20}t_{5.5}^{-9/10},\,\,\,\,\,\,\,\,\,
\end{eqnarray} 
and the transition time for the system to be optically thin to GeV gamma rays is 
\begin{equation}
t_{\tau_{\rm BH}=1}\simeq5.3\times{10}^3~{\rm s}~{(\tilde{Z}+1)}^{10/9}\mu_e^{-10/9}D_{*,-2}^{11/9}{\mathcal E}_{\rm ej,51}^{-1/2}M_{\rm ej,1}^{7/18}. 
\label{eq:ttBH}
\end{equation}
As inferred in Fig.~\ref{fig:EMLC} left, we see the rising of GeV light curves until $\sim t_{\tau_{\rm BH}=1}$ for $D_*\gtrsim0.1$. Without CCSM, as indicated by the thin curves, the system is optically thin to the Bethe-Heitler pair production process just after the breakout from a progenitor. 
For gamma-ray spectra shown in Fig.~\ref{fig:EMsp}, the Bethe-Heitler attenuation is no longer important.  
Even with CCSM, given that CR acceleration at $\tau_T\lesssim c/V_s$, the Bethe-Heitler pair production can be important only at earlier times, and it is negligible for high-velocity shocks with $V_s\gtrsim10^4~{\rm km}~{\rm s}^{-1}$.  

\begin{figure*}[tb]
\includegraphics[width=0.325\linewidth]{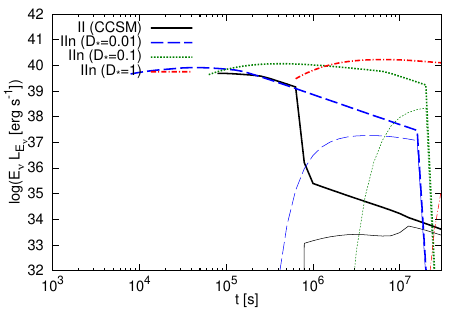}
\includegraphics[width=0.325\linewidth]{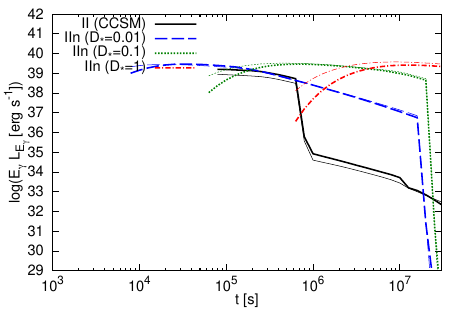}
\includegraphics[width=0.325\linewidth]{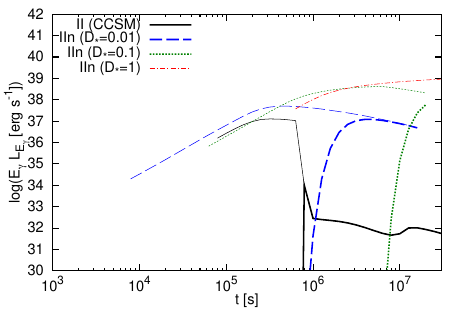}
\caption{Left panel: light curves of high-energy gamma rays at $\varepsilon_\gamma=1$~GeV (thick) and $\varepsilon_\gamma=100$~GeV (thin). Middle panel: light curves of soft gamma rays at $\varepsilon_\gamma=1$~MeV (thick) and hard x rays at $\varepsilon_\gamma=10$~keV (thin). Right panel: light curves of radio waves at $\nu=100$~GHz with (thick) and without (thin) external matter attenuation. In all three panels, we include $L_{\rm ph}$ for CCSM-interacting SNe II, in which $D_*=0.01$ is used.
\label{fig:EMLC2}
}
\end{figure*}

Note that only sufficiently high-energy gamma rays interact with SN photons, and the two-photon annihilation becomes important at 
\begin{eqnarray}
\tilde{\varepsilon}_{\gamma\gamma-{\rm sn}}\approx \frac{m_e^2c^4}{\varepsilon_{\rm sn}}\simeq260~{\rm GeV}~{(\varepsilon_{\rm sn}/1~{\rm eV})}^{-1},
\end{eqnarray}
where $\varepsilon_{\rm sn}\sim3k{\mathcal T}_{\rm sn}$. 
The optical depth to $\gamma\gamma\rightarrow e^+e^-$ is estimated to be
\begin{eqnarray}
\tau_{\gamma\gamma}\approx\frac{3}{16}\sigma_{\gamma\gamma}n_\gamma R_s&\simeq&8200~{\rm max}[1,\tau_T]L_{\rm sn,42.5}{(\varepsilon_{\rm sn}/1~{\rm eV})}^{-1} \nonumber\\
&\times&D_{*,-2}^{1/10}{\mathcal E}_{\rm ej,51}^{-9/10}M_{\rm ej,1}^{7/20}t_{5.5}^{-9/10}, \,\,\,\,\, \,\,\,\,\,
\end{eqnarray}
where $n_\gamma$ is the photon number density. The interaction time, $t_{\gamma\gamma}$, is shown in Fig.~\ref{fig:timescale}. The ratio of the light crossing time to the interaction time has a local maximum around $\tilde{\varepsilon}_{\gamma\gamma-{\rm sn}}$, and the corresponding break/cutoff energy is expected to be $\sim10-100$~GeV below $\tilde{\varepsilon}_{\gamma\gamma-{\rm sn}}$. 
The transition time for the system to be optically thin for TeV gamma rays would occur much later around 
\begin{eqnarray}
t_{\tau_{\gamma\gamma}=1}&\simeq&7.0\times{10}^9~{\rm s}~L_{\rm sn,42.5}^{9/10}{(\varepsilon_{\rm sn}/1~{\rm eV})}^{-9/10}\nonumber\\
&\times&D_{*,-2}^{1/9}{\mathcal E}_{\rm ej,51}^{-1/2}M_{\rm ej,1}^{7/18},
\end{eqnarray} 
which suggests that gamma rays at TeV and higher energies are effectively absorbed inside and outside the system. This can be clearly seen in gamma-ray spectra shown in Fig.~\ref{fig:EMsp}.  
In Fig.~\ref{fig:EMLC2} left, we show light curves of 100~GeV gamma rays (thin curves). For SNe II with CCSM, such very high-energy gamma-ray emission is suppressed during the main interaction phase and escapes after the shock enters the ordinary RSG wind medium. Light curves for SNe IIn with different values of $D_*$ are also shown in Fig.~\ref{fig:EMLC2}. One sees that the 100~GeV gamma-ray fluxes at early times anticorrelate with $D$, implying that observations within appropriate time windows are critical to utilize very high-energy gamma rays as a promising probe of SNe IIn. 
Note that the two-photon annihilation with x-ray photons from thermal bremsstrahlung is not significant in our cases. However, for SNe Ibc, x-ray photons from shock breakout can provide important target photons for $\gamma\gamma\rightarrow e^+e^-$ as well as photomeson production, as shown in Ref.~\cite{Kashiyama:2012zn}.  

Analytically, the pionic gamma-ray flux at GeV energies obeys the following scaling,
\begin{equation}
E_\gamma F_{E_\gamma}^{\pi^0}
\propto
e^{-\tau_{\rm BH}(t)}
\left\{
\begin{array}{lr} 
\varepsilon_\gamma^{2-s_{\rm cr}}t^{-0.3} & \mbox{($t < t_{f_{pp}=1}$)} \\
\varepsilon_\gamma^{2.1-s_{\rm cr}}t^{-1.1} & \mbox{($t > t_{f_{pp}=1}$)} 
\end{array} 
\right., 
\label{eq:gammascaling}
\end{equation}
which is similar to Eq.~(\ref{eq:nuscaling}) but with matter attenuation at early times (see Fig.~\ref{fig:EMLC} left). 

\subsection{Inverse-Compton (cascade) emission}
%
\begin{figure}[tb]
\includegraphics[width=\linewidth]{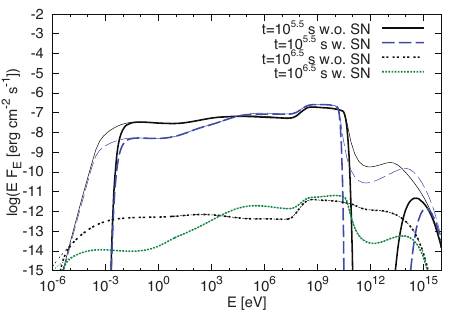}
\caption{Energy fluxes of nonthermal electromagnetic emission from a Galactic SN II (CCSM) for $D_*=0.01$ at $d=10$~kpc. Thick/thin curves represent spectra with/without external matter attenuation, and $s_{\rm cr}=2.2$ is used. One sees that synchrotron emission is suppressed in the presence of ordinary SN emission with $L_{\rm ph}$ from the photosphere.  
\label{fig:cascade}
}
\end{figure}

Secondary electrons and positrons as well as primary electrons lose their energies via various cooling processes. In our setup, there are three characteristic injection energies. The hadronic injection occurs through the decay of charged pions and the two-photon annihilation process. In the former case, the injection Lorentz factor of electrons and positrons produced via $pp$ interactions is
\begin{equation}
\gamma_{e,h}\approx \frac{m_\pi}{4m_e}\simeq68.
\end{equation}
In the latter case, the injection Lorentz factor of electrons and positrons is
\begin{equation}
\gamma_{e,\gamma\gamma}\approx \frac{\tilde{\varepsilon}_{\gamma\gamma-{\rm sn}}}{2m_ec^2}\simeq2.5\times{10}^5~{(\varepsilon_{\rm sn}/1~{\rm eV})}^{-1}.
\end{equation}
In addition, leptonic injection occurs through primary electron acceleration. Observations of SN remnants and CR spectra on Earth suggest $\epsilon_e\sim{10}^{-4}-10^{-3}$, and it has been shown that the hadronic component is dominant when $\epsilon_e\lesssim0.2\epsilon_{\rm cr,-1}$ in the calorimetric limit~\cite{Murase:2013kda}. Although the leptonic injection is available in {\textsc AMES}, it is negligible during the main interaction phase for CCSM parameters considered in this work. 
  
The radiative cooling Lorentz factor of electrons and positrons is given by 
\begin{eqnarray}
\gamma_{e,\rm rc} &\approx&\frac{6\pi m_e c}{\sigma_T B^2(1+Y)t}\nonumber\\
&\simeq&1.6~\varepsilon_{B,-2}^{-1}D_{*,-2}^{-1}t_{5.5}{(1+Y)}^{-1}
\end{eqnarray}
where $Y$ is the Compton Y parameter, 
\begin{eqnarray}
Y&\approx&\frac{-1+\frac{U_{\rm sn}}{U_B}+\sqrt{{\left(1+\frac{U_{\rm sn}}{U_B}\right)}^2+\frac{2}{3}{\rm min}[f_{pp},1]\frac{\epsilon_{\rm cr}V_s}{\varepsilon_Bc}}}{2}\nonumber\\ 
&\rightarrow& \frac{U_{\rm sn}}{U_B} \,\,\,\,\, \mbox{(for $\epsilon_{\rm rad} \gg \epsilon_{\rm cr}$)} 
\end{eqnarray}
in the Thomson limit, where $U_{\rm sn}$ is the energy density of SN thermal photons in the emission region. Different from the pulsar-driven SN scenario, where the SN emission itself is attributed to the regeneration of nonthermal synchrotron radiation~\cite{Murase:2014bfa}, we expect that the Compton Y parameter is governed by seed photons originating from the thermal radiation. 
If the SN emission is powered by the CSM interaction, as in SNe IIn, we obtain
\begin{eqnarray}
Y&\simeq&0.57~{\rm max}[1,\tau_T]\epsilon_{\rm rad,-0.6}\varepsilon_{B,-2}^{-1}D_{*,-2}^{-1/10}\nonumber\\
&\times&{\mathcal E}_{\rm ej,51}^{9/20}M_{\rm ej,1}^{-7/20}t_{5.5}^{-1/10}, 
\end{eqnarray}
while for $L_{\rm sn}\approx L_{\rm ph}$ we have
\begin{eqnarray}
Y\approx\frac{{\rm max}[1,\tau_T]L_{\rm ph}V_s}{\varepsilon_B L_{s}c}
&\simeq&7.2~{\rm max}[1,\tau_T]L_{\rm ph,42.5} \varepsilon_{B,-2}^{-1}\nonumber\\
&\times&D_{*,-2}^{-4/5}{\mathcal E}_{\rm ej,51}^{-9/10}M_{\rm ej,1}^{7/10}t_{5.5}^{1/5}.\,\,\,\,\,
\end{eqnarray}
Cooling timescales of electrons and positrons at $t_{\nu{\rm max}}$ are shown in Fig.~\ref{fig:timescale}. One finds $\gamma_{e,\rm rc} < \gamma_{e,h}, \gamma_{e,\gamma\gamma}$, implying that the system is in the fast-cooling regime for leptons. Ignoring Coulomb losses, the transition time from the fast to slow cooling regimes is
\begin{equation}
t_{f\rightarrow s}\simeq 1.3\times{10}^{7}~{\rm s}~\varepsilon_{B,-2}D_{*,-2}(1+Y). 
\end{equation}
Thus, we may expect the fast cooling in most of the interaction duration $t_{\rm end}$ especially if $Y\gg1$. 
In the Thomson limit, the ratio of synchrotron to IC cooling timescales is $t_{\rm syn}/t_{\rm IC} \approx U_{\rm sn}/U_B$, and the analytical estimates agree with the numerical results well. In the limit that SN thermal emission in the optical band is powered by CSM interaction, the synchrotron cooling is comparable to the IC cooling (see Fig.~\ref{fig:timescale} left and right). On the other hand, as expected in ordinary SNe II-P, the IC cooling is dominant in the presence of SN optical photons from the photosphere (see Fig.~\ref{fig:timescale} middle). We also note that the Klein-Nishina effect is important for leptons with $\gtrsim10-100$~GeV energies. 

However, there is one complication from the standard fast-cooling spectrum. Owing to strong Coulomb losses in dense CSM, the lepton distribution can be modified and the cooling Lorentz factor does not have to be $\gamma_{e,\rm rc}$.
By comparing Coulomb losses with radiative losses, the ``Coulomb break'' Lorentz factor in the fast-cooling regime can be introduced as
\begin{eqnarray}
\gamma_{e,\rm Cou} &\sim&{\left(\frac{270\pi m_e c^2 n_e}{B^2(1+Y)}\right)}^{1/2}\nonumber\\
&\simeq&84~\varepsilon_{B,-2}^{-1/2}\mu_e^{-1/2}D_{*,-2}^{1/10}{\mathcal E}_{\rm ej,51}^{-9/20}\nonumber\\
&\times&M_{\rm ej,1}^{7/20}t_{5.5}^{1/10}{(1+Y)}^{-1/2}.\,\,\,\,\,\,\,\,\,\,
\label{eq:gammacou}
\end{eqnarray}
This analytical estimate agrees with numerical results shown in Fig.~\ref{fig:timescale}, and Coulomb losses are important in SNe II with CCSM. Note that this Coulomb break Lorentz factor is different from that in the slow cooling regime~\cite{Uchiyama:2002gn}, where it is defined by comparing the Coulomb cooling time with the dynamical time. The transition to the standard fast-cooling regime occurs at
\begin{eqnarray}
t_{\gamma_{e,\rm rc}=\gamma_{e,\rm Cou}}&\simeq&2.5\times{10}^{7}~{\rm s}~\varepsilon_{B,-2}^{5/9}\mu_e^{-5/9}D_{*,-2}^{11/9}\nonumber\\
&\times&{\mathcal E}_{\rm ej,51}^{-9/18}M_{\rm ej,1}^{7/18}{(1+Y)}^{5/9}.
\end{eqnarray}

In the fast-cooling limit, the spectrum of electrons and positrons in the quasisteady state is roughly expressed as
\begin{eqnarray}
n_{\gamma_e}^e
\propto \gamma_e^{-q}
\propto
\left\{
\begin{array}{lr} 
\gamma_e^{-2+\Delta s} & \mbox{($\gamma_e< \gamma_{e,h}$)} \\
\gamma_e^{-\tilde{s}-1+\Delta s} & \mbox{($\gamma_{e,h}\leq \gamma_e$)} 
\end{array} 
\right., 
\end{eqnarray}
where $\tilde{s}$ is the effective injection index, which is $\tilde{s}\approx s$ for secondary injection from $pp$ interactions. If there is an additional contribution from electromagnetic cascades through $\gamma\gamma\rightarrow e^+e^-$, the energy spectrum is flatter, which may lead to $\tilde{s}\sim2$. 
Here $\Delta s$ represents the possible spectral hardening effect due to Coulomb losses. If the Coulomb cooling is more important than the radiative cooling, i.e., $\gamma_e<\gamma_{e,\rm Cou}$, we have $\Delta s\approx2$. As a result, most of the energy comes from leptons with $\gamma_{e,b}={\rm max}[\gamma_{e,h},\gamma_{e,\rm Cou}]$, and their characteristic energy of IC emission from secondaries injected through $pp$ interactions is
\begin{eqnarray}
\varepsilon_{\rm IC}^b\approx2\gamma_{e,b}^2\varepsilon_{\rm sn}&\simeq&9.2~{\rm keV}~(\varepsilon_{\rm sn}/1~{\rm eV}){(\gamma_{e,b}/\gamma_{e,h})}^2,\,\,\,
\end{eqnarray}
which is expected in the hard x-ray range. For the $\gamma\gamma\rightarrow e^+e^-$ injection component, we obtain
\begin{eqnarray}
\varepsilon_{\rm IC}^{\gamma\gamma}\approx2\gamma_{e,\gamma\gamma}^2\varepsilon_{\rm sn}&\simeq&130~{\rm GeV}~{(\varepsilon_{\rm sn}/1~{\rm eV})}^{-1},
\end{eqnarray}
which is expected in the gamma-ray range. 

Assuming the fast-cooling regime, the resulting IC luminosity from CR-induced electrons and positrons is
\begin{eqnarray}
\varepsilon_\gamma L_{\varepsilon_\gamma}&\approx&\frac{Y}{2(1+Y)}\frac{f_{\rm sup}^{\rm mat}}{6}{\rm min}[f_{pp},1]\frac{\epsilon_{\rm cr} L_s}{{\mathcal R}_{\rm cr10}}g_{\rm Cou}{\left(\frac{\varepsilon_\gamma}{\varepsilon_{\rm IC}^b}\right)}^{2-\beta_{\rm IC}}\nonumber\\
&\simeq&8.4\times{10}^{38}~{\rm erg}~{\rm s}^{-1}~{(Y/[1+Y])}g_{\rm Cou}\nonumber\\
&\times&{\rm min}[f_{pp},1]f_\Omega\epsilon_{\rm cr,-1}{\mathcal R}_{\rm cr10,1}^{-1}{\left(\frac{\varepsilon_\gamma}{\varepsilon_{\rm IC}^b}\right)}^{2-\beta_{\rm IC}}\nonumber\\
&\times&D_{*,-2}^{7/10}{\mathcal E}_{\rm ej,51}^{27/20}M_{\rm ej,1}^{-21/20}t_{5.5}^{-3/10}e^{-\tau_{\rm Comp}-\tau_{\rm pe}},
\label{eq:IClum}
\end{eqnarray}
where $\beta_{\rm IC}=3/2$ or $\tilde{s}/2$ or $2/3$ for $\varepsilon_{\rm IC}<\varepsilon_{\rm IC}^b$ and $\beta_{\rm IC}=(\tilde{s}+2)/2$ for $\varepsilon_{\rm IC}^b\leq\varepsilon_{\rm IC}$, respectively, and $g_{\rm Cou}={(\varepsilon_\gamma^b/\varepsilon_\gamma^h)}^{(2-\tilde{s})/2}$.
Here $f_{\rm sup}^{\rm mat}$ is the suppression fraction of x rays, which is dominated by Compton scattering and photoelectric absorption in the unshocked CSM. 
As shown in Fig.~\ref{fig:EMsp}, the energy spectrum of the CR-induced IC emission is flat in $E_\gamma F_{E_\gamma}$ especially above $E_{\rm IC}^b$, which is consistent with an analytical expectation by Eq.~(\ref{eq:IClum}).   

In order to demonstrate the impacts of SN emission from the photosphere, Fig.~\ref{fig:cascade} shows the results with $L_{\rm ph}$. As analytically expected by considering $Y\gg1$, additional external photons enhance the IC cascade emission especially at late times, whereas the energy flux below keV energies is smaller because the synchrotron flux significantly contributes at these energies (see also next subsection). 
The electromagnetic cascade makes overall energy spectra very flat, and because of $\gamma_{e,\rm Cou}\sim\gamma_{e,h}$ in our cases the effects of Coulomb losses are hardly seen in the resulting electromagnetic spectra. 
The IC cascade spectrum is extended from the x-ray band to the gamma-ray energy range, being overwhelmed by the pionic gamma-ray component above 0.1~GeV, as seen in Figs.~\ref{fig:EMsp} and \ref{fig:cascade}. Although the break from $\pi^0$ decay should exist, the pionic gamma-ray component may stand out by only a factor of  $\sim3-5$. We also note that thermal bremsstrahlung emission is dominant in the $1-100$~keV band (cf. Fig.~\ref{fig:thspectrum}), so higher energies especially in the MeV--GeV bands are better to detect the IC cascade component. 

At MeV energies, where the nonthermal component is likely to be dominant, photons lose their energies mainly via Compton downscatterings, and the Compton optical depth is estimated to be
\begin{equation}
\tau_{\rm Comp}\simeq0.025~(\tilde{Z}+1)\mu_e^{-1}D_{*,-2}^{11/10}{\mathcal E}_{\rm ej,51}^{-9/20}M_{\rm ej,1}^{7/20}t_{5.5}^{-9/10},
\end{equation} 
and the transition time for the system to be optically thin is 
\begin{equation}
t_{\tau_{\rm Comp}=1}\simeq5.3\times{10}^3~{\rm s}~{(\tilde{Z}+1)}^{10/9}\mu_e^{-10/9}D_{*,-2}^{11/9}{\mathcal E}_{\rm ej,51}^{-1/2}M_{\rm ej,1}^{7/18}. 
\label{eq:ttComp}
\end{equation} 

Given that hadronic components from $pp$ interactions are dominant, the CR-induced IC energy flux follows 
\begin{eqnarray}
E_\gamma F_{E_\gamma}^{\rm IC}
&\propto&
e^{-\tau_{\rm Comp}(t)}Y(t){(1+Y(t))}^{-1}\\\nonumber
&\times&
\left\{
\begin{array}{lr} 
\varepsilon_\gamma^{2-\beta_{\rm IC}}\varepsilon_{\rm sn}^{\beta_{\rm IC}-2}t^{-0.3} & \mbox{($t < t_{f_{pp}=1}$)} \\
\varepsilon_\gamma^{2-\beta_{\rm IC}}\varepsilon_{\rm sn}^{\beta_{\rm IC}-2}t^{-1.1} & \mbox{($t > t_{f_{pp}=1}$)} 
\end{array} 
\right., 
\label{eq:icscaling}
\end{eqnarray}
which agrees with the numerical curves shown in the middle panels of Figs.~\ref{fig:EMLC} and \ref{fig:EMLC2}.

\subsection{Synchrotron (cascade) emission}
Primary electrons and secondary electron-positron pairs emit synchrotron emission while they are advected in the shock downstream. Sufficiently high-energy pionic gamma rays create electron-positron pairs, which also contribute to additional synchrotron emission though cascades.  
When the hadronuclear component is dominant, their characteristic frequencies are given by
\begin{eqnarray}
\nu_{\rm syn}^b&\approx&\frac{3}{4\pi}\gamma_{e,b}^2\frac{eB}{m_ec}\nonumber\\
&\simeq&760~{\rm GHz}~\varepsilon_{B,-2}^{1/2}D_{*,-2}^{1/2}t_{5.5}^{-1}{(\gamma_{e,b}/\gamma_{e,h})}^2,
\end{eqnarray}
and
\begin{eqnarray}
\varepsilon_{\rm syn}^{\gamma\gamma}&\approx&\frac{3}{4\pi}\gamma_{e,{\gamma\gamma}}^2\frac{heB}{m_ec}\nonumber\\
&\simeq&44~{\rm keV}~{(\varepsilon_{\rm sn}/1~{\rm eV})}^{-2}\varepsilon_{B,-2}^{1/2}D_{*,-2}^{1/2}t_{5.5}^{-1},
\end{eqnarray}
respectively. Reference~\cite{Murase:2013kda} pointed out that secondary synchrotron emission peaks in the high-frequency radio (submillimter or millimeter) band, which provides a smoking gun of ion acceleration in interaction-powered SNe or Type IIn SNe (see also Refs.~\cite{Petropoulou:2016zar,Matsuoka:2019djv}). 
However, for more ordinary SNe, we find that the CR-induced synchrotron emission can be suppressed by two effects. First, as discussed in the previous subsection, IC cooling can be more efficient due to external SN photons, which is especially the case for Type II-P SNe. The second is Coulomb scattering with electrons, through which energy of nonthermal leptons can rather be used for plasma heating.  

Assuming the fast-cooling regime, the differential synchrotron luminosity is analytically expressed as
\begin{eqnarray}
\nu L_\nu^{\rm syn}&\approx&\frac{1}{2(1+Y)}\frac{f_{\rm sup}}{6}{\rm min}[f_{pp},1]\frac{\epsilon_{\rm cr} L_s}{{\mathcal R}_{\rm cr10}}
g_{\rm Cou}{\left(\frac{\nu}{\nu_b}\right)}^{2-
\beta_{\rm syn}}\nonumber\\
&\simeq&8.4\times{10}^{38}~{\rm erg}~{\rm s}^{-1}~{(1+Y)}^{-1}g_{\rm Cou}\nonumber\\
&\times&{\rm min}[f_{pp},1]f_\Omega\epsilon_{\rm cr,-1}{\mathcal R}_{\rm cr10,1}^{-1}{\left(\frac{\nu}{\nu_{\rm syn}^b}\right)}^{2-\beta_{\rm syn}}\nonumber\\
&\times&D_{*,-2}^{7/10}{\mathcal E}_{\rm ej,51}^{27/20}M_{\rm ej,1}^{-21/20}t_{5.5}^{-3/10}\frac{e^{-\tau_{\rm ff}}}{1+\tau_{\rm sa}},
\label{eq:synlum}
\end{eqnarray}
where $\beta_{\rm syn}=3/2$ or $\tilde{s}/2$ or or $2/3$ for $\nu_{\rm syn}<\nu_{\rm syn}^b$ and $\beta_{\rm syn}=(\tilde{s}+2)/2$ for $\nu_{\rm syn}\geq\nu_{\rm syn}^b$, respectively, and $f_{\rm sup}$ is the suppression factor that mainly consists of SSA and free-free absorption. In Figs.~\ref{fig:EMsp} and \ref{fig:cascade}, low-energy spectra originate from CR-induced synchrotron emission. Energy fluxes of radio emission from secondary pairs are comparable to those of CR-induced IC emission for interaction-powered SNe with $L_{\rm sn}\sim L_{s}$ because of $Y\lesssim1$. However, for SNe with $L_{\rm sn}\sim L_{\rm ph}\gg L_{s}$, we find that radio emission can be strongly suppressed by IC cooling, as clearly indicated in Fig.~\ref{fig:cascade}.
Note that in the last expression of Eq.~(\ref{eq:synlum}) we have assumed that the emission region is completely screened or embedded by the CSM. In reality, the CCSM is clumpy or aspherical, in which a significant fraction of radio emission may escape. The radio light curves without $f_{\rm sup}^{\rm mat}$ are shown in the right panels of Figs.~\ref{fig:EMLC} and \ref{fig:EMLC2}. The spectra without $f_{\rm sup}^{\rm mat}$ are also shown in Figs.~\ref{fig:EMsp} and \ref{fig:cascade} with the thin curves. The hadronic scenario predicts that the synchrotron emission below $\nu_{\rm syb}^b$ has a hard spectrum with $\beta_{\rm syn}\lesssim3/2$, which is rather robust thanks to Coulomb losses that make the spectra harder. If a softer spectrum is observed in the radio band, the existence of primary electrons or additional processes such as reacceleration via turbulence may be inferred. 

The SSA optical depth is analytically estimated by~\citep{Murase:2013kda}
\begin{equation}
\tau_{\rm sa}(\nu)={\eta_{\rm sa}}\frac{3 \tilde{f}_e L_s e}{4\pi R_s V_s m_p c^2 B\gamma_n^5}{\left(\frac{\nu}{\nu_n}\right)}^{-(q+4)/2}
\end{equation}
where $\tilde{f}_e\approx(1/6){\rm min}[1,f_{pp}]\epsilon_{\rm cr}/{\mathcal R}_{\rm cr10}/(\tilde{s}-1)\equiv 0.045\tilde{f}_{e,*}$ for the hadronic scenario, $\gamma_{e,n}\approx{\rm min}[\gamma_{e,h},\gamma_{e,c}]$, $\nu_n=3\gamma_{e,n}^2eB/(4\pi m_ec)$, and $\eta_{\rm sa}$ is
\begin{equation}
\eta_{\rm sa}=\tilde{\xi}_q\frac{\pi^{\frac{3}{2}}2^{\frac{q}{2}}}{3^{\frac{3}{2}}}\frac{\Gamma(\frac{q}{4}+\frac{11}{6})\Gamma(\frac{q}{4}+\frac{1}{6})\Gamma(\frac{q}{4}+\frac{3}{2})}{\Gamma(\frac{q}{4}+2)},
\end{equation}
where $\tilde{\xi}_q=(q-1)$. Although the SSA optical depth can be affected by Coulomb losses, as noted above, Eq.~(\ref{eq:gammacou}) suggests that $\gamma_{e,\rm Cou}$ is not too far from $\gamma_{e,h}$ and considering $\gamma_{e,\rm rc}<\gamma_e<\gamma_{e,h}$ would give a conservative result. With $q=2$, the SSA optical depth at $\nu_{\rm syn}^h$ is approximated to be
\begin{equation}
\tau_{\rm sa}(\nu_{\rm syn}^h)\simeq0.020~\tilde{f}_{e,*}\varepsilon_{B,-2}^{-3/2}D_{*,-2}^{-3/5}{\mathcal E}_{\rm ej,51}^{9/20}M_{\rm ej,1}^{-7/20}t_{5.5}^{9/10}{(1+Y)}^{-1},
\end{equation}
where we note that $\tilde{f}_{e,*}$ depends on $t$. Then, for $\nu_{\rm sa}<\nu_{\rm syn}^{h}$, we obtain
\begin{equation}
\nu_{\rm sa}\simeq210~{\rm GHz}~\tilde{f}_{e,*}^{1/3}D_{*,-2}^{3/10}{\mathcal E}_{\rm ej,51}^{3/20}M_{\rm ej,1}^{-7/60}t_{5.5}^{-21/30}{(1+Y)}^{-1/3},
\end{equation}
and the transition time when the emission region is optically thin to SSA is
\begin{eqnarray}
t_{\tau_{\rm sa}=1}&\simeq&1.7\times{10}^5~{\rm s}~\tilde{f}_{e,*}^{10/21}D_{*,-2}^{3/7}{\mathcal E}_{\rm ej,51}^{3/14}M_{\rm ej,1}^{-1/6}\nonumber\\
&\times&{(1+Y)}^{-10/21}\nu_{11.5}^{-10/7}.
\label{eq:ttsa}
\end{eqnarray}

On the other hand, the free-free optical depth in the unshocked CSM is approximated to be
\begin{equation}
\nu_{\rm ff}\simeq2.2~{\rm THz}~\tilde{Z}^{10/21}\mu_e^{-20/21}{\mathcal T}_{\rm ext,5}^{-9/14}D_{*,-2}^{23/21}{\mathcal E}_{\rm ej,51}^{-9/14}M_{\rm ej,1}^{1/2}t_{5.5}^{-9/7},
\end{equation}
in the fully ionized limit. This gives the radio breakout time against the free-free absorption process,
\begin{eqnarray}
t_{\tau_{\rm ff}=1}&\simeq&1.4\times{10}^6~{\rm s}~\tilde{Z}^{10/27}\mu_e^{-20/27}{\mathcal T}_{\rm cs,5}^{-1/2}
\nonumber\\
&\times&D_{*,-2}^{23/27}{\mathcal E}_{\rm ej,51}^{-1/2}M_{\rm ej,1}^{7/18}
\nu_{11.5}^{-7/9}.
\label{eq:ttff}
\end{eqnarray}
These results imply that the free-free absorption is more severe than the SSA absorption in cases of the simplest spherical CSM. However, we note that the radio escape fraction can be readily larger if the CSM is aspherical or clumpy, and/or if the ionization fraction is lower in the unshocked CSM. 

Finally, the CR-induced synchrotron flux from secondaries produced via inelastic $pp$ interactions follows
\begin{eqnarray}
E_\gamma F_{E_\gamma}^{\rm syn}
&\propto&
e^{-\tau_{\rm ff}(t)}{[(1+Y(t))(1+\tau_{\rm sa}(t))]}^{-1}\\\nonumber
&\times&
\left\{
\begin{array}{lr} 
\nu^{2-\beta_{\rm syn}} t^{\beta_{\rm syn}-2.3} & \mbox{($t < t_{f_{pp}=1}$)} \\
\nu^{2-\beta_{\rm syn}} t^{\beta_{\rm syn}-3.1} & \mbox{($t > t_{f_{pp}=1}$)} 
\end{array} 
\right.,
\label{eq:synscaling}
\end{eqnarray}
which agrees with the numerical curves shown in Fig.~\ref{fig:EMLC2} right. The cases without $\tau_{\rm ff}$ are also shown in the right panels of Figs.~\ref{fig:EMLC} and \ref{fig:EMLC2}. 
The results imply that submillimeter or millimeter emission from CCSM-interacting SNe II is promising only around $t_{\rm end}$ unless the escape fraction is larger than $f_{\rm sup}^{\rm mat}$.

\section{Implications}
\subsection{Multiwavelength (gamma--x--radio) relation and testability of the hadronic scenario}
%
\begin{figure}[tb]
\includegraphics[width=\linewidth]{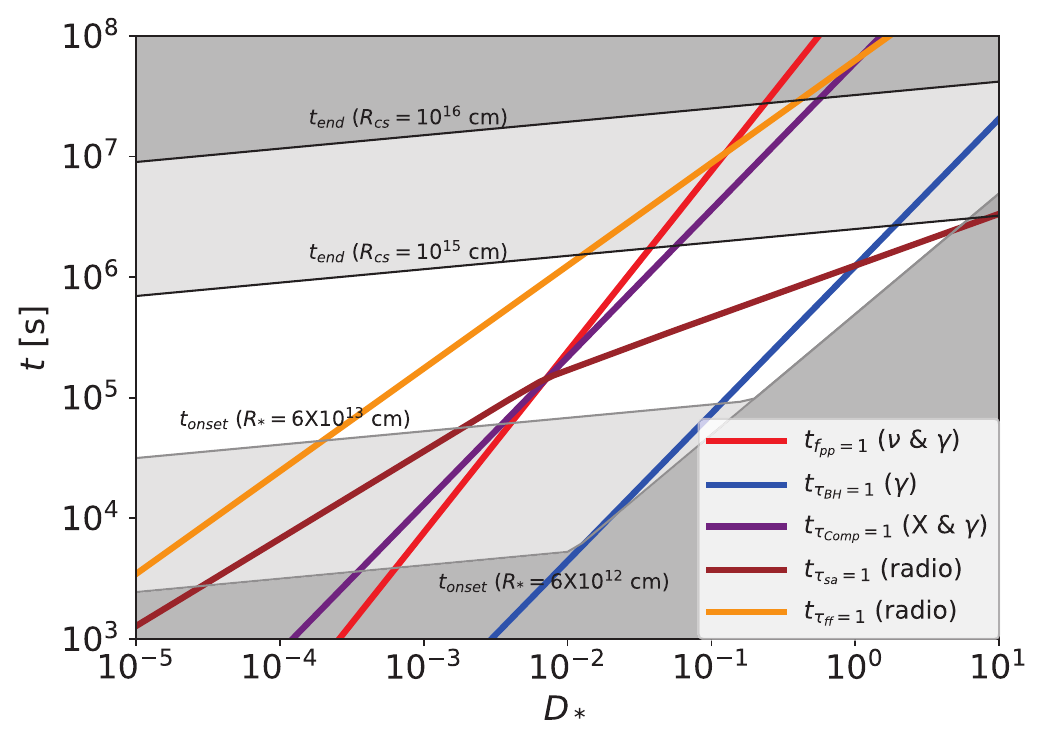}
\caption{Characteristic time windows for observations of neutrinos, gamma rays, x rays and radio waves for Type II SNe interacting with CCSM. Strong CSM interaction accompanied by CR acceleration occurs from $t_{\rm onset}$ to $t_{\rm end}$. See text for details. 
\label{fig:typicaltime}
}
\end{figure}
\begin{figure*}[tb]
\includegraphics[width=0.325\linewidth]{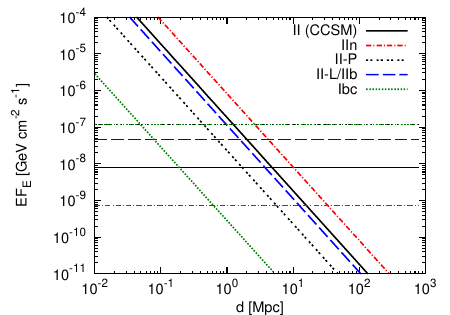}
\includegraphics[width=0.325\linewidth]{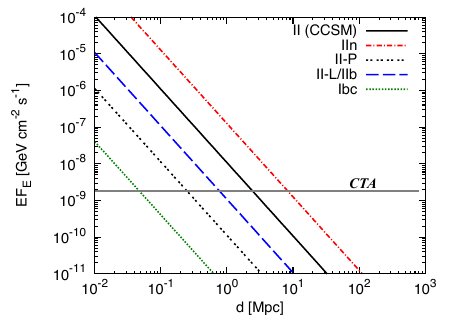}
\includegraphics[width=0.325\linewidth]{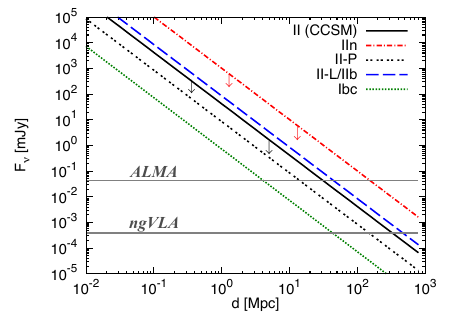}
\caption{Detection horizons of hadronic gamma rays at $E_\gamma=1$~GeV with {\it Fermi}-LAT sensitivities (left panel) and $E_\gamma=50$~GeV with the CTA sensitivity (middle panel). The right panel shows detection horizons of hadronic radio emission at 100~GHz with ALMA and ngVLA sensitivities, where upper limits are shown for SNe II (CCSM) and SNe IIn by ignoring external matter attenuation. Note that for CCSM SNe II $D_*=0.01$ is used without $L_{\rm ph}$. 
\label{fig:detectionhorizon}
}
\end{figure*}

\textsc{AMES} allows us to consistently calculate gamma-ray, x-ray and radio emissions, considering electromagnetic cascades. 
In addition to the ``multimessenger'' connection shown in Eq.~(\ref{eq:gammalum}), one may expect the following ``multiwavelength'' relation among nonthermal luminosities:
\begin{equation}
\frac{\varepsilon_\gamma L_{\varepsilon_\gamma}^{\pi^0}}{{\left(\frac{\varepsilon_\gamma}{0.8~{\rm GeV}}\right)}^{2-s}}
\sim\frac{4(1+Y)\nu L_{\nu}^{\rm IC}}{Y{\left(\frac{\varepsilon_\gamma}{\varepsilon_{\rm IC}^b}\right)}^{2-\beta_{\rm IC}}}
\sim\frac{4(1+Y)\nu L_{\nu}^{\rm syn}}{{\left(\frac{\nu}{\nu_{\rm syn}^b}\right)}^{2-\beta_{\rm syn}}}, 
\label{eq:mwrelation}
\end{equation}
which essentially reflects the ratio between $\pi^\pm$ and $\pi^0$ produced via inelastic $pp$ interactions. 
Observationally, deviations from Eq.~(\ref{eq:mwrelation}) can be caused by contributions from other components by e.g., thermal bremsstrahlung and nonthermal emission from primary electrons. 

Simultaneous observations are important for testing the above relation. The next Galactic SN would be ideal for detailed studies, and in the previous sections the results on $EF_E$ were presented for a SNe at $d=10$~kpc. 
On the other hand, Galactic SNe are rare and the core-collapse SN rate in the Milky Way is estimated to be $\sim3~{\rm century}^{-1}$~\cite{Adams:2013ana}. The rates of Type II-P, II-L/IIb and IIn SNe are $\sim50$\% and $\sim15-20$\%, and $\sim10$\%, respectively~\cite{Smith:2010vz,Li:2010kc}. Thus, searching for extragalactic SNe is relevant especially for rarer types of SNe including Type IIn SNe. 
Then, it would be useful to clarify characteristic time windows, e.g., for neutrino-triggered optical followup observations that have been suggested for extragalactic SNe~\cite{Murase:2006mm,Yoshida:2022idr,Pitik:2023vcg}. 

In Fig.~\ref{fig:typicaltime}, we show several timescales that may represent characteristic observational time windows. 
For neutrinos which can leave the system without attenuation, $t_{f_{pp}=1}$, which corresponds to the peak time of $t L_\nu$ or neutrino fluences, is shown [see Eq.~(\ref{eq:tfpp})].  
For electromagnetic emission, we also consider various breakout times when the system is optically thin to relevant attenuation or scattering processes. 
For pionic gamma-ray and IC cascade components, the Bethe-Heitler pair production [Eq.~(\ref{eq:tfpp})] and Compton [Eq.~(\ref{eq:tfpp})] processes are relevant as an effective attenuation process, respectively. 
For radio waves, the SSA [Eq.~(\ref{eq:ttsa})] and free-free [Eq.~(\ref{eq:ttff})] absorption are considered at $\nu=10^{2.5}$~GHz.
In addition, the range from $t_{\rm onset}$ to $t_{\rm end}$ for SNe II with CCSM is depicted. 
For x rays and gamma rays, we find that $t_{f_{pp}=1}$ is longer than the breakout timescales in the range of $D_*\gtrsim0.01$. This implies that the characteristic time windows are essentially governed by $t_{f_{pp}=1}$. For radio waves, if the free-free absorption process is critical, $t_{\tau_{ff}=1}$ is the most relevant timescale for radio followup observations. However, if the free-free attenuation is reduced by e.g., aspherical or clumpy CSM, the characteristic time for hadronic radio emission may be either $t_{f_{pp}=1}$ or $t_{\tau_{\rm sa}=1}$. 

\begin{figure*}[tb]
\includegraphics[width=0.325\linewidth]{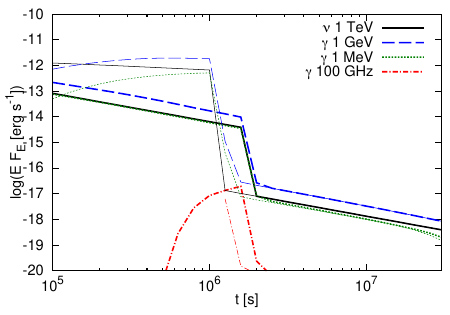}
\includegraphics[width=0.325\linewidth]{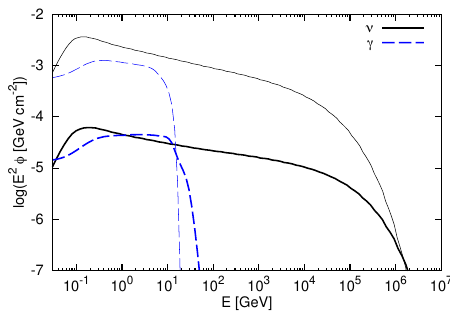}
\includegraphics[width=0.325\linewidth]{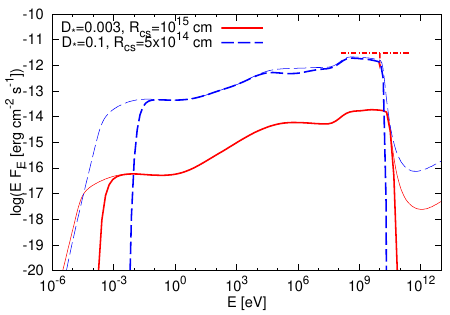}
\caption{Left panel: light curves of high-energy neutrinos at $E_\nu=1$~TeV, high-energy gamma rays at $E_\gamma=1$~GeV, soft gamma rays at $E_\gamma=1$~MeV, and radio waves at 100~GHz, for SN 2023ixf with $D_*=0.003$ and $R_{\rm cs}={10}^{15}$~cm (thick) and $D_*=0.1$ and $R_{\rm cs}=5\times{10}^{14}$~cm (thin). 
Middle panel: energy fluences of $\nu_e+\bar{\nu}_e+\nu_\mu+\bar{\nu}_\mu+\nu_\tau+\bar{\nu}_\tau$ and cascaded gamma rays, integrated over $\approx30$~d, from SN 2023ixf, where $D_*=0.003$ and $R_{\rm cs}={10}^{15}$~cm (thick) and $D_*=0.1$ and $R_{\rm cs}=5\times{10}^{14}$~cm (thin) are considered. 
Right panel: energy fluxes of nonthermal electromagnetic emission (at $t=10^6$~s) with (thick) and without (thin) external matter attenuation. The preliminary gamma-ray upper limit by {\it Fermi}-LAT is overlaid. In all three panels, $d=6.9$~Mpc and $s_{\rm cr}=2.2$ are used, and $L_{\rm ph}$ is included based on Ref.~\cite{Hiramatsu:2023inb}.
\label{fig:2023ixf}
}
\end{figure*}

The detectability of high-energy neutrinos and gamma rays from extragalactic interacting SNe has been studied (see Refs.~\cite{Petropoulou:2016zar,Murase:2018okz} for the earliest works). 
In Fig.~\ref{fig:detectionhorizon} we presented detection horizons for GeV gamma rays (left), sub-TeV gamma rays (middle), and radio emission (right).
For the detectability of GeV gamma rays by {\it Fermi}, both the energy flux and {\it Fermi} sensitivity for each SN type are shown for the optimized observation time, which is found by maximizing the ratio of the signal flux to the sensitivity. One sees that detection of pionic gamma rays is promising for not only Galactic SNe but also SNe at the Small Megellanic Cloud at 60~kpc and the Large Magellanic Cloud at 50~kpc. In particular, SNe II with CCSM are detectable up to $d\sim5$~Mpc for $D_*\sim0.01$ (SN 2013fs-like). See also Fig.~1 of Ref.~\cite{Kheirandish:2022eox} for the list of nearby galaxies (see also Ref.~\cite{Valtonen-Mattila:2022nej}). 
For the detectability of 50~GeV gamma rays by imaging atmospheric Cherenkov telescopes (IACTs), we assume the Cherenkov Telescope Array (CTA) north array with an integration time of 50~hr~\cite{Hassan+17}. The energy flux for each SN type is shown for the optimized observational time that is often the peak time caused by gamma-ray suppression via the two-photon annihilation process. 
Even for SNe II with CCSM and IIn, the detection is possible up to $d\sim3-10$~Mpc because the suppression due to $\gamma\gamma\rightarrow e^+e^-$ is more important for brighter SNe including SNe with stronger interaction with CCSM. We also note that future MeV gamma-ray satellites such as {\it AMEGO-X}~\cite{Caputo:2022xpx} and {\it e-ASTROGAM}~\cite{e-ASTROGAM:2016bph} can play important roles in detecting CR-induced IC cascade emission. For example, the {\it e-ASTROGAM} sensitivity for $10^6$~s is $\sim{10}^{-11}~{\rm erg}~{\rm cm}^{-2}~{\rm s}^{-1}$ at 1~MeV, which allows us to hadronic MeV gamma-ray emission from SNe II with CCSM at $d\lesssim2-3$~Mpc. 
The detectability of radio emission is strongly affected by free-free absorption even in the high-frequency band at $\sim100$~GHz. For CCSM-interacting SNe II and IIn, we indicate the cases without the free-free absorption as upper limits, and the integration time is assumed to be 300~s for ALMA (Atacama Large Millimeter/submillimeter Array)~\cite{ALMA} and 1~hr for ngVLA (next-generation Very Large Array)~\cite{ngVLA}, respectively.  
One sees that CR-induced synchrotron emission can be detected more easily than pionic gamma rays, although nondetection of radio signals is not surprising for SNe with dense CSM. Note that for SNe with ordinary CSM (i.e., SNe II-P, II-L/IIb, and Ibc without CCSM) the radio detectability is even conservative because it is enhanced by synchrotron emission from primary electrons.

\subsection{Application to SN 2023ixf}
A nearby Type II SN observed in the galaxy M101 at $d=6.9$~Mpc~\cite{Jacobson-Galan:2023ohh} provided unique opportunities to investigate the interaction with CCSM. 
Early spectroscopic observations suggest $\dot{M}_w\sim10^{-3}-10^{-2}~M_\odot~{\rm yr}^{-1}$ and $R_{\rm cs}\sim(0.5-1)\times{10}^{15}$~cm~\cite{Jacobson-Galan:2023ohh,Bostroem:2023dvn,Teja:2023hcm}. Although modeling of light curves may favor large values of $\dot{M}_w$~\cite{Hiramatsu:2023inb}, we adopt $D_*=0.1$ and $R_{\rm cs}=5\times{10}^{14}$~cm~\cite{Jacobson-Galan:2023ohh} as the case motivated by optical observations. 
Detailed observations indicate that the CSM structure is complex, and the CSM profile may be steeper around $R_{\rm cs}\sim{\rm a~few}\times{10}^{14}$~cm~\cite{Zimmerman:2023mls,Li:2023vux}.
This is supported by x-ray observations, whose spectrum is consistent with thermal bremsstrahlung emission with a photoelectric absorption feature~\cite{Grefenstette:2023dka,Chandra:2023kda}, and we also adopt $D_*=0.003$ and $R_{\rm cs}=10^{15}$~cm.
The apparent discrepancies among different observations may suggest that the CCSM is not spherical, which has been inferred by high-resolution spectroscopy and polarization measurements~\cite{Smith:2023kqo,Vasylyev:2023mtl}. 

In Fig.~\ref{fig:2023ixf} left, we show neutrino and photon light curves for these two parameter sets of $D_*$ and $R_{\rm cs}$. 
For $D_*=0.1$, the system is mostly calorimetric during the interaction with CCSM, and the light curves of neutrinos and gamma rays are nearly flat although gamma rays at early times are subject to Compton or Bethe-Heitler attenuation. For $D_*=0.003$, the light curves decline more rapidly because of $f_{pp}<1$, and the gamma-ray attenuation is negligible from MeV to GeV energies.  
Radio emission is suppressed mainly due to the free-free absorption process especially for $D_*=0.1$, but it will break out at $t\sim 10-20$~d when the shock reaches the CSM edge. The resulting synchrotron emission may explain the VLA detection of radio waves around that time~\cite{VLASN2023}. 
We also expect that the primary leptonic component (not shown in Fig.~\ref{fig:2023ixf}) dominates late-time radio emission.

Neutrino and gamma-ray fluences are presented in Fig.~\ref{fig:2023ixf} middle. As shown in Ref.~\cite{Kheirandish:2022eox}, the predicted numbers of throughgoing muon events for $s_{\rm cr}=2.0$ are approximately 0.0007 and 0.02 events for $D_*=0.003$ and $D_*=0.1$, respectively. Thus, the CCSM-interaction scenario is consistent with the null results reported by the IceCube fast response analysis~\citep{IceCubeSN2023} (see also Ref.~\cite{Guetta:2023mls}).
Neutrino and photon fluxes at 30~d after the explosion are also shown in Fig.~\ref{fig:2023ixf} right, where one sees that CR-induced synchrotron emission is strongly suppressed due to optical SN emission from the photosphere. Instead, IC (cascade) emission is enhanced especially in the MeV range although in the x-ray range it is still below the thermal bremsstrahlung flux. Furthermore, one sees that the predicted gamma-ray fluxes for $\epsilon_{\rm cr}\sim0.01-0.1$ are consistent with the preliminary upper limit reported by {\it Fermi}~\cite{FermiSN2023}, which leads to the constraint on the CR energy fraction, $\epsilon_{\rm cr}\lesssim0.2-200$, reflecting the range of $D_*$. This limit is complementary to that of SN 2010jl, $\epsilon_{\rm cr}\lesssim0.05-0.1$~\cite{Murase:2018okz}, in that the results are obtained for Type II SNe that are more ordinary than SNe IIn. 
Because ordinary SNe II (not SNe IIn) at $d\sim7$~Mpc would still be too far to detect neutrino and gamma-ray signals, these results strengthen the importance of not missing nearby SNe at $d\lesssim1-5$~Mpc. The core-collapse SN rate in local galaxies is higher than in the rate estimated from the star-formation rate in the continuum limit~\cite{Ando:2005ka,Nakamura:2016kkl}, and it can be as high as $\sim0.8~{\rm yr}^{-1}$ for SNe within $\sim4-5$~Mpc. 
Such nearby SNe are promising targets for not only gamma-ray detectors such as {\it Fermi} and IACTs but also the high-energy neutrino detector network~\cite{Kheirandish:2022eox} and Hyper-Kamiokande~\cite{Hyper-Kamiokande:2018ofw}.

\section{Summary}\label{sec:summary}
We investigated high-energy multimessenger emission from interacting SNe for different classes of SNe. 
Considering the velocity distribution of SNe ejecta, we presented detailed results on broadband multimessenger emission, including light curves and spectra. The system can be nearly calorimetric for inelastic $pp$ collisions, leading to promising GeV-TeV signals of high-energy neutrinos and pionic gamma rays although TeV or higher-energy gamma rays may be significantly affected by the two-photon annihilation process. Electromagnetic cascades are developed in the presence of Coulomb losses, in which both IC and synchrotron components are relevant in the x-ray/soft gamma-ray and radio range, respectively. We also derived the analytical expressions that describe electromagnetic fluxes from the interacting SNe, which are consistent with the results of numerical calculations.    
The overall energy spectrum is expected to be flat over wide energy ranges, especially from MeV to $\sim10$~GeV energies (although the low-energy break from neutral pion decay would still be visible). Hadronic synchrotron emission predicts hard spectra with $\beta_{\rm syn}\lesssim3/2$ in the radio band, and Coulomb losses can make the spectra harder. The synchrotron fluxes may also be significantly suppressed by efficient IC cooling due to SN photons, and various absorption processes affect the spectra in the radio band.  

The hadronic scenario predicts the unique multimessenger relationship, which can be tested with simultaneous observations of nearby SNe. The next Galactic SN would be the most promising event that would enable us to confront the theory with observational data. However, detection of nonthermal signals from extragalactic SNe may be challenging, and we discussed characteristic time windows. For high-energy neutrinos and gamma rays, the time at which the system is effectively transparent to inelastic $pp$ interactions is typically the most important. The gamma-ray signal is detectable up to a few Mpc for ordinary SNe II, although Type IIn SNe may be observed up to dozens of Mpc. TeV or higher-energy gamma rays would be significantly absorbed by SN photons, and detection with $\sim10-50$~GeV energies is relevant for IACTs such as HESS, MAGIC, VERITAS and CTA. 
For radio waves and soft x rays, attenuation or scattering processes affect the detectability of interacting signatures. Hadronic radio emission is predicted to be dominate over leptonic radio emission during the interaction with dense CCSM, but the radio signal at early times is suppressed by SSA and free-free absorption so that the detectability depends on details of the CSM structure. If the escape fraction is $\sim1$\% due to clumpy or aspherical geometry, high-frequency radio emission from SNe with CCSM may be detectable up to dozens of Mpc. 
We also discussed SN 2023ixf as an example of interest. Nondetection of neutrinos and gamma rays is consistent with our theoretical predictions. While CR-induced IC emission in the x-ray range is overwhelmed by thermal bremsstrahlung emission, CR-induced synchrotron emission may significantly contribute to the observed radio emission. It is very important not to miss future nearby SNe within $\sim3-5$~Mpc, and all-sky monitoring of the brightest SNe with followup observations (by e.g., ASAS-SN, Tomo-e Gozen and ZTF) are necessary. 

The method used in {\textsc AMES} for calculating multimessenger spectra of interacting SNe can be applied to arbitrary evolution of $R_s$ and $V_s$ as well as energy distributions of CRs and external SN radiation fields. Given that {\sc AMES} includes the photomeson production and photodisintegration processes~\cite{Zhang:2023ewt}, the SN module can be used for not only the SN reverse shock but also other transients powered by shock interactions such as winds from compact binary mergers~\cite{Murase:2017snw} and tidal disruption events~\cite{Murase:2020lnu}. The results with a combination of the Complete History of Interaction-Powered Supernovae ({\textsc CHIPS})~\cite{Takei:2021ize} will be presented in the forthcoming paper.  

\medskip
\begin{acknowledgments}
The multimessenger spectral templates presented in this work and M18 are available on Github \cite{Github}. The public version of the SN module in \textsc{AMES} will be available upon request. 

K.M. thanks Kunihito Ioka, Daichi Tsuna, and B. Theodore Zhang for useful discussions. 
The work of K.M. is supported by the NSF Grants No.~AST-1908689, No.~AST-2108466, and No.~AST-2108467, and KAKENHI No.~20H01901 and No.~20H05852. 
\end{acknowledgments}


\bibliography{kmurase.bib}

\end{document}